\documentclass[journal]{IEEEtran}

\usepackage{graphicx}
\usepackage[colorlinks]{hyperref}
\hypersetup{
	colorlinks=true, 
	linkcolor=blue!70!black, 
	citecolor=red!70!black, 
	filecolor=blue!70!black, 
	urlcolor=blue!70!black, 
}
\usepackage{amsfonts,amssymb,amsbsy,amsmath}
\usepackage{amsxtra}
\usepackage{amsthm}
\usepackage{amssymb}
\usepackage{multirow}
\usepackage{mathabx}
\usepackage{bm}
\usepackage{steinmetz}
\usepackage{chngcntr}
\usepackage[dvipsnames]{xcolor} 
\usepackage{tikz}
\usepackage{longtable}
\usepackage{supertabular}
\usepackage{optidef}
\usepackage{threeparttable}
\usepackage{pgfplots}
\usepackage{booktabs}
\usepackage{xurl}

\usepackage{styling}
\usepackage{grffile}

\graphicspath{{Figures/}}

\ifCLASSINFOpdf
\else
\fi
\ifCLASSOPTIONcompsoc
  \usepackage[caption=false,font=normalsize,labelfont=sf,textfont=sf]{subfig}
\else
  \usepackage[caption=false,font=footnotesize,labelformat=simple]{subfig}
\fi

\begin{document}

\newcommand{\red}[1]{\textcolor{PairedF}{#1}}
\newcommand{\blue}[1]{\textcolor{PairedB}{#1}}
\newcommand{\green}[1]{\textcolor{PairedD}{#1}}
\newcommand{\orange}[1]{\textcolor{PairedG}{#1}}

\newcommand{\mut}{\vec{u}_{\theta}}
\newcommand{\mup}{\vec{u}_{\varphi}}
\newcommand{\mua}{\vec{u}_{\alpha_l}}

\newcommand{\SDD}{\bm{S}^{\mathrm{DD}}}
\newcommand{\SDP}{\bm{S}^{\mathrm{DP}}}
\newcommand{\SPD}{\bm{S}^{\mathrm{PD}}}
\newcommand{\SPP}{\bm{S}^{\mathrm{PP}}}

\newcommand{\aD}{a^{\mathrm{D}}}
\newcommand{\bD}{b^{\mathrm{D}}}
\newcommand{\aP}{a^{\mathrm{P}}}
\newcommand{\bP}{b^{\mathrm{P}}}
\newcommand{\maD}{\bm{a}^{\mathrm{D}}}
\newcommand{\mbD}{\bm{b}^{\mathrm{D}}}
\newcommand{\maP}{\bm{a}^{\mathrm{P}}}
\newcommand{\mbP}{\bm{b}^{\mathrm{P}}}

\newcommand{\mED}{\bm{E}^{\mathrm{D}}}
\newcommand{\mEP}{\bm{E}^{\mathrm{P}}}
\newcommand{\meD}{\bm{e}^{\mathrm{D}}}
\newcommand{\meP}{\bm{e}^{\mathrm{P}}}
\newcommand{\eD}{e^{\mathrm{D}}}
\newcommand{\eP}{e^{\mathrm{P}}}
\newcommand{\newe}{\hat{e}}
\newcommand{\newme}{\hat{\bm{e}}}
\newcommand{\newmE}{\hat{\bm{E}}}

\newcommand{\rc}{r}
\newcommand{\mr}{\bm{r}}
\newcommand{\mR}{\bm{R}}
\newcommand{\xnr}{\rc_n^{\mathrm{Re}}}
\newcommand{\xni}{\rc_n^{\mathrm{Im}}}

\newcommand{\NP}{M}
\newcommand{\ND}{N}

\newcommand{\ret}{\operatorname{ret}}
\newcommand{\diag}{\operatorname{diag}}
\newcommand{\eig}{\operatorname{eig}}
\newcommand{\st}{\mathrm{subject \, to}}
\newcommand{\trace}{\operatorname{tr}}
\newcommand{\vect}{\operatorname{vec}}
\newcommand{\ddiag}{\operatorname{Diag}}

\newcommand{\ji}{\mathrm{j}}
\newcommand{\e}{\mathrm{e}}

\renewcommand{\Re}{\mathrm{Re}}
\renewcommand{\Im}{\mathrm{Im}}

%
\title{Optimization of Embedded Element Patterns of Reactively Loaded Antenna Arrays}


\author{Albert~Salmi, Miloslav~Capek, ~\IEEEmembership{Senior~Member,~IEEE}, Lukas~Jelinek, Anu~Lehtovuori, and~Ville~Viikari,~\IEEEmembership{Senior~Member,~IEEE}
\thanks{Manuscript received August 16, 2024; revised \today. This work was supported by the Czech Science Foundation under project~\mbox{No.~21-19025M} and by the Business Finland through the RF Sampo Project. The work of Albert Salmi was supported in part by Helsingin Puhelinyhdistys (HPY) Research Foundation.}
\thanks{A. Salmi, A. Lehtovuori, and V. Viikari are with the Department of Electronics and Nanoengineering, Aalto University, Espoo, P.O.\ Box 15500, 00076 Aalto, Finland, (e-mail: albert.salmi@aalto.fi; anu.lehtovuori@aalto.fi; ville.viikari@aalto.fi).}
\thanks{M. Capek and L. Jelinek are with the Department of Electromagnetic
Field, Faculty of Electrical Engineering, Czech Technical University in Prague,
166 27 Prague, Czech Republic (e-mail: miloslav.capek@fel.cvut.cz; lukas.
jelinek@fel.cvut.cz)}
\thanks{This article has supplementary downloadable material available at \url{https://doi.org/10.5281/zenodo.14412065}, provided by the authors. This includes MATLAB codes, simulation files, and result data for reproducing the presented results.}}
\maketitle

\begin{abstract}
This paper introduces a framework for synthesizing reactively loaded antennas and antenna arrays. The framework comprises two main components: computing the fundamental bound using the semi-definite relaxation and finding a realizable solution via optimization on a Riemannian manifold. The embedded element patterns are subject to the optimization with two distinct goals under study: focusing the radiation in a single direction or synthesizing patterns with desired shapes. The reactive terminations of passive antenna elements serve as optimization variables. We demonstrate the framework using a connected bowtie-slot antenna and antenna array with both beam-focusing and beam-shaping targets. The tests show that the optimization on the Riemannian manifold yields superior results compared to existing methods, such as the genetic algorithm. This is particularly evident in the most complex and extensive problem, which requires the synthesis of shaped embedded element patterns for a sparse reactively loaded antenna array with a limited field of view.
\end{abstract}


%
\IEEEpeerreviewmaketitle

\section{Introduction}

The synthesis of antenna radiation patterns is a critical task in the field of antenna engineering, particularly concerning applications in satellites, mobile base stations, and radars, where the transmission and reception of radio waves in specific directions are necessary \cite{Encinar2011, Su2022, Yang2019, Puskely2022, skobelev2011book}. The emergence of reconfigurable intelligent surfaces (RIS) has further emphasized the need for precise control of electromagnetic scattering \cite{Tirkkonen2024, Wu2020}. In the design of sparse antenna arrays, the pattern synthesis is required to limit the field of view to mitigate grating lobes~\cite{Maximidis2020, Salmi2024, skobelev2011book}.

Designing an antenna with a specific radiation pattern poses a significant challenge. Although synthesizing an electric current density distribution meeting a desired radiation pattern is feasible, realizing the optimal current in available space is often extremely difficult \cite{Bouwkamp1945, Jelinek2017, Kormilainen2020}. A common approach is to discretize the available space to smaller antenna elements, and construct the desired radiation pattern as a linear combination of the elements' radiation patterns \cite{Laue2018, Kormilainen2021, Cai2019, Petrolati2014, Skobelev1998}. While various methods have been developed to compute the feeding coefficients of the elements, a common issue arises from the need for a feeding network for the elements, which can consume a substantial amount of space or prove impractical to implement.

A popular way to circumvent the need for feeding networks and increase freedom of the design is through the use of passively loaded aperture-coupled elements, wherein the driven antenna element couples to \cite{Harrington1972, Corcoles2015, Lamey2021, Maximidis2020, Maximidis2022, Sun2004, Fezai2013, Fezai2015, Georgiadis2022, Jiang2022, Salmi2024}. These passive elements can be parasitic antenna elements with passive loads in the ports, or, generally, any topologies presenting a load-terminated port in the structure, such as gaps in pixel layout~\cite{Jiang2022}. The passive elements scatter the coupled waves, with their terminations defining the phase and magnitude of the scattered waves. By tuning the terminations, the radiation pattern of the antenna can be manipulated. Reactive loads are commonly used in terminating the passive elements because power dissipation in the loads is then avoided.

Various methods exist for optimizing these reactively loaded antennas in specific use cases, with the most extensively researched scenario being the reactively controlled antenna, where only one element is driven and the radiation is tuned to a single direction by adjusting the loads of the passive elements \cite{Corcoles2015, Nyffenegger2022, Lang2018, Harrington1978, Fezai2013, Fezai2015, Georgiadis2022}. This type of antenna is commonly known as an electronically steerable parasitic array radiator (ESPAR) \cite{Sun2004, Junwei2005}. A more intricate optimization challenge involves terminating the passive elements to achieve an antenna radiation pattern of a predetermined shape. 

Furthermore, passive elements can be integrated into antenna arrays, where multiple elements are driven \cite{Corcoles2015, Lamey2021, Maximidis2020, Maximidis2022, Lang2017}. Reactively loaded antenna arrays present two distinct optimization challenges: directing the radiation to a single direction or shaping the embedded element patterns (EEPs) of the driven elements. Shaping the EEPs allows for the limitation of the antenna array's field of view to suppress grating lobes~\cite{Salmi2024_pca, Salmi2024, skobelev2011book, Maximidis2020}. However, the synthesis of EEPs using static, reactively terminated passive elements represents an extremely challenging task due to the NP-hardness of the constant-modulus-constrained optimization problem~\cite{Cao2017} and currently lacks scalable and robust optimization methods.

We present an optimization-based framework for computing the reactive terminations. It can be applied in both single and multi-driven antennas with both beam-focusing and beam-shaping targets. The framework consists of two optimization methods. The first method gives a tight fundamental bound for the optimization problem and is based on semi-definite relaxation \cite{Boyd1996_sdp, Luo2010, Wang2020, Fuchs2014}. The second finds a realization and is based on optimization on a Riemannian manifold and uses the Riemannian augmented Lagrangian method (RALM) \cite{absil2008_book, boumal2023_book, Liu2020}. These methods are compared with another semi-definite relaxation-based approach, presented in \cite{Corcoles2015}, and the genetic algorithm. The novelty compared to similar works \cite{Corcoles2015, Salmi2024, Salmi2024_pca} is that the individual EEPs are subject to the optimization which enables beam-steering functionality with fixed reactive loads and without amplitude tapering in the feeding coefficients of the driven elements.

The rest of this paper is organized as follows. Section~\ref{sec:method} presents the mathematical model of a reactively loaded antenna array, outlines the optimization objectives for the four problem types, and introduces the novel optimization methods of the framework. The mathematical formulations of the methods are expressed in the appendices. Section ~\ref{sec:demonstration} demonstrates the optimization methods using a connected bowtie-slot antenna array as a test antenna, while the results are discussed in Section~\ref{sec:discussion}, and Section~\ref{sec:conclusion} draws conclusions. All codes producing the presented results, as well as the simulation models, are available as supplementary material.

\section{Optimization of reactively loaded antennas}
\label{sec:method}
This section introduces the mathematical model of a general reactively loaded antenna array. The array comprises multiple driven antenna elements and passive elements terminated with passive loads. The ports of the passive elements are referred to as scatterer ports. The mathematical model captures the embedded element patterns of the driven elements as a function of the loads connected to the scatterer ports. 

\begin{figure}[t]
    \centering
    \def\svgwidth{1\linewidth}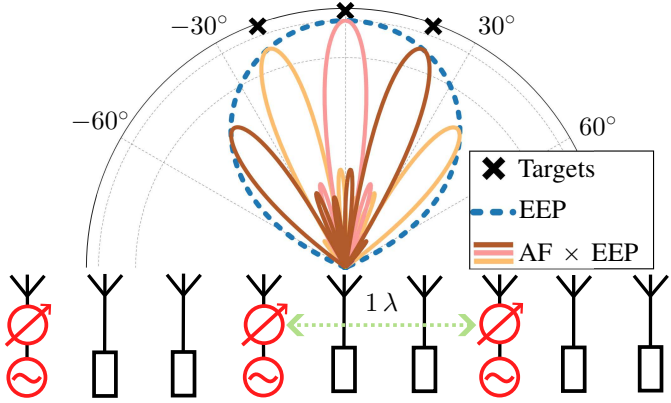%
    \caption{Optimization problem illustrated assuming that all EEPs are identical. The solid lines illustrate the antenna array's radiation patterns using the array factor (AF) when the beam is steered in three directions. The driven elements are connected to the sources (depicted by the red color), and the remaining elements are terminated by fixed passive loads.}
    \label{fig:idea_figure}
\end{figure}

The concept of a sparse reactively loaded antenna array is illustrated in Fig.~\ref{fig:idea_figure}, which shows an example with three driven elements and six passive elements. In this example, the inter-element spacing between the driven elements is set to one wavelength, $\lambda$. Consequently, the beam-steering range must be limited to avoid grating lobes \cite{kildal2015, skobelev2011book}. The optimization problem is to determine a fixed set of scatterer port terminations that maximizes beam-steering gain toward specified target directions by appropriately phasing the driven elements. 

Following the formulation of the mathematical model, the optimization problems are defined and categorized. These problems are classified into four categories based on the complexity arising from the number of driven antenna elements and the number of beam directions considered. Finally, two optimization approaches are presented: a semi-definite relaxation-based approach for determining the fundamental performance bounds, and the Riemannian augmented Lagrangian method for identifying local solutions to the non-convex problem.

\subsection{Model of reactively loaded antenna array}
\label{sec:method_model}
\begin{figure}[t]
    \centering
    \includegraphics[width=1\linewidth]{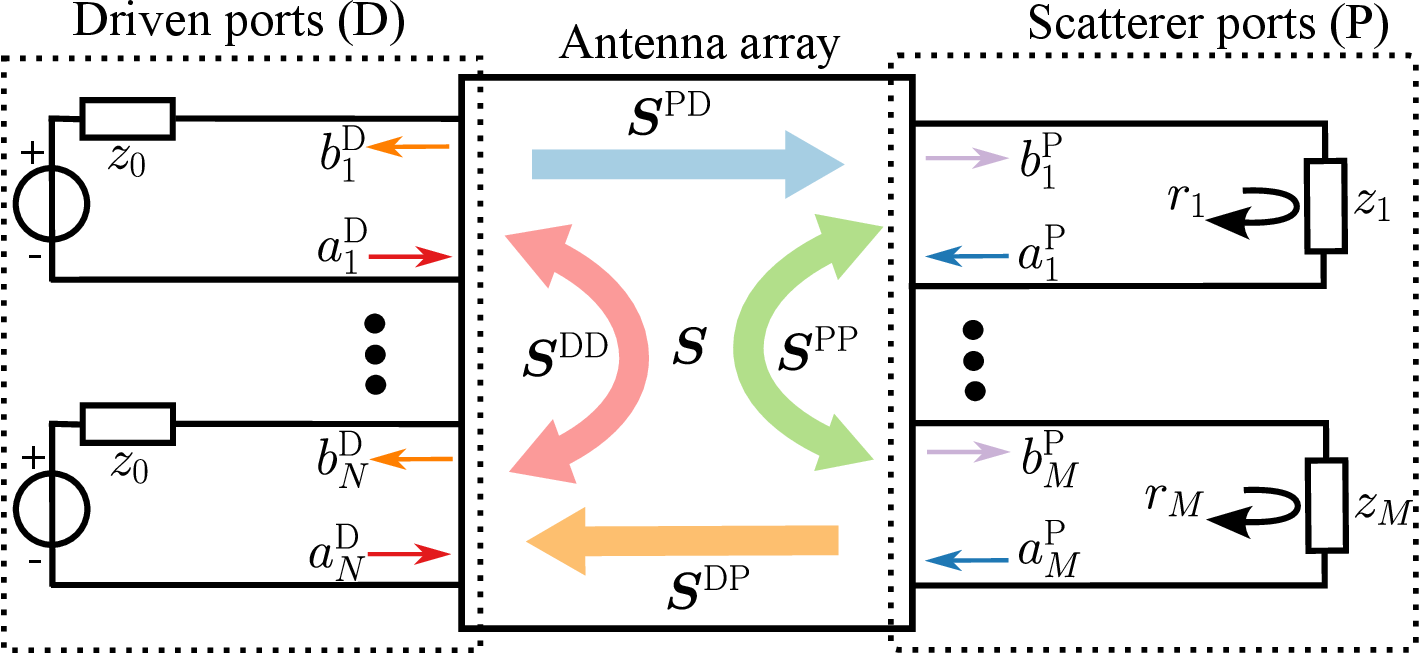}
    \caption{N-port model of a passively loaded antenna array.}
    \label{fig:nport}
\end{figure}
In the proposed optimization methods, EEPs alone are subject to optimization, not the final array patterns as is done in \cite{Corcoles2015}, for instance. This approach is chosen because the final array pattern depends on the EEPs, and the feeding coefficients of the driven elements can be computed optimally after the EEPs are optimized \cite{Salmi2023, Salmi2024}. In addition, optimizing the EEPs consequently optimizes the realized gain, which also considers impedance matching, assuming that the generator impedances are fixed. As compared to a classical array design, our approach enables the realization of the feeding coefficients for optimal beamforming because the impedance level is optimized through the EEPs. Therefore, we derive a function describing the EEPs of the driven elements as functions of passive element terminations. The derivation is based on \cite{Salmi2024}.

A typical problem setting could be the following: \textit{Design an antenna array with the given beam-steering range radiating the given polarization.} The two polarization components of the EEPs are analyzed separately. In this work, we focus on problems with a single specified polarization, although the dual-polarized operation could be applied similarly as is done in \cite{Salmi2024}.

Passively loaded antenna arrays can be modeled using S-parameters, considering the ratios of incident and scattered waves \cite{Salmi2024, Jiang2022, Lamey2021, Maximidis2020, Fezai2013} or Z-parameters, considering the ratios between voltages and currents \cite{Nyffenegger2022, Lang2018, Lang2017, Lang2017MTT, Kormilainen2021}. The two representations are fundamentally equivalent \cite{Pozar2012}. We use the S-parameter formulation because the magnitudes of the S-parameters are limited between 0 and 1, whereas Z-parameters can approach to infinity, provoking significant computational challenges.

The passively loaded antenna array is initially modeled as an N-port in which all the ports are considered to be actively driven. Using electromagnetic simulation software, the S-matrix and the EEPs related to each port of the N-port are computed. After the initial simulation, $\ND$ ports are considered driven, and the rest $\NP$ are assigned passively terminated, resulting in the N-port in Fig.~\ref{fig:nport}.

The scattering matrix~$\bm{S}$, computed in the initial simulation, is partitioned to blocks~$\SDD$, $\SDP = (\SPD)^{\mathrm{T}}$, and $\SPP$, describing the coupling between the driven ports, between the driven and scatterer ports, and between the scatterer ports, respectively. Using the N-port in Fig.~\ref{fig:nport} and the block matrix notation, the ratios can be expressed as
\begin{align}
    \bm{b} = \begin{bmatrix}
        \mbD \\
        \mbP
    \end{bmatrix}
    =
    \begin{bmatrix}
        \SDD & \SDP \\
        \SPD & \SPP
    \end{bmatrix}
    \begin{bmatrix}
        \maD \\
        \maP
    \end{bmatrix}
    =
    \bm{S} \bm{a},
\end{align}
where $\bm{a}$ and $\bm{b}$ are the vectors of incident and scattered waves, respectively.

To compute the EEP of a driven port~$n$ after terminating the scatterer ports to loads~$z_1, \ldots, z_{\NP}$, the driven port~$n$ is excited with unit magnitude and zero phase. That is, $\maD = \bm{u}_n$, where $\bm{u}_n = [0 \ldots 1 \ldots 0]^\mathrm{T}$ with one at entry $n$. The vector of incident waves to the scatterer ports is then \mbox{$\maP_n \in \mathbb{C}^{\NP}$}, and it can be computed as
\begin{align} \label{eq:maPn}
    \maP_n = (\mR^{-1} - \SPP)^{-1} \SPD \bm{u}_n.
\end{align}
The terminations of the scatterer ports are introduced in the diagonal matrix $\mR$,
\begin{align}
    \mR = \operatorname{diag}(\mr) = \begin{bmatrix}
        \rc_1 & \hdots & 0 \\
        \vdots & \ddots & \vdots \\
        0 & \hdots & \rc_{\NP}
    \end{bmatrix},
\end{align}
whose elements are the reflection coefficients at the scatterer ports with respect to the generator impedances, that is,
\begin{align}
    \label{eq:r_vs_z}
    \rc_m = \frac{z_m - z_0}{z_m + z_0}.
\end{align}
The reflection coefficients are the variables under optimization, as shown in the following sections.

An EEP related to a port~$i$, that could be a driven or scatterer port, is denoted as~$\vec{E}_i(\theta, \varphi)$. A co-polarized component of an EEP toward a direction $(\theta_l, \varphi_l)$, is 
\begin{align}
    e_{il} = \vec{E}_i(\theta_l, \varphi_l) \cdot \vec{u}^{\mathrm{co}}(\theta_l, \varphi_l).
\end{align}
The unit vector $\vec{u}^{\mathrm{co}}(\theta_l, \varphi_l)$ defines the co-polarization.

When the scatterer ports are terminated to loads, the EEP of the driven element $n$ toward $(\theta_l, \varphi_l)$ is
\begin{align}
    \newe_{nl} = \eD_{nl} + (\maP_n)^{\mathrm{T}} \meP_l,
    \label{eq:new_enl}
\end{align}
where $\eD_{nl}$ is the EEP of the driven element $n$ in the case where both driven and scatterer ports are terminated to generator impedances, and $\meP_l \in \mathbb{C}^{\NP}$ is a vector of scatterer ports' EEPs toward the direction $l$ \cite{Salmi2024}.

\subsection{Optimization objectives}
\label{sec:method_objectives}
Table~\ref{tab:problem_types} summarizes the addressed design problems and compares their complexities. Firstly, the optimization objectives are divided into single-beam and multi-beam objectives. In the single-beam case ($L=1$), we consider only one direction $(\theta_0, \varphi_0)$ toward which we strive to minimize or maximize the EEPs or realize given magnitudes for the EEPs. In the multi-beam cases, the EEPs are analyzed at multiple directions $\{(\theta_l, \varphi_l)\}_{l=1}^L$. The single and multi-beam goals are also discussed as beam focusing and beam shaping goals, respectively.

The second dimension of the optimization problem size is the number of driven elements, $\ND$. If we design a single-fed or periodic antenna array using unit-cell simulations, $\ND = 1$. When designing antenna arrays using full-array simulation, that is, without approximation of the array being infinitely large, then $\ND$ is larger than one. Also, there might be multiple feeding points inside an element, for example, to achieve dual-polarized operation. The third dimension would be frequency. However, in this work, we limit the analysis to a single frequency point only.

\begin{table}[t]
    \centering
    \caption{Complexities and proposed solutions for the four considered problem types: single-driven single-beam (SDSB), multi-driven single-beam (MDSB), single-driven multi-beam (SDMB), multi-driven multi-beam (MDMB).}
    \begin{tabular}{ccccc}
        Problem type & Feeds & Beams & Solutions & Complexity \\ \toprule
        \addlinespace[0.3em] SDSB & 1 & 1 & \cite{Lang2018, Harrington1978, Fezai2015, Georgiadis2022} & \green{simple}\\ 
        MDSB & $\ND$ & 1 &  \cite{Corcoles2015, Nyffenegger2022} & \orange{moderate} \\ 
        SDMB & 1 & $L$ & \cite{Maximidis2020_symposium, Maximidis2020_1} & \orange{moderate} \\ 
        MDMB & $\ND$ & $L$ & \cite{Salmi2024, Maximidis2020, Lamey2021} & \red{difficult} \\ \bottomrule
    \end{tabular}
    \label{tab:problem_types}
\end{table}

The four problem types are labeled as single-driven single-beam (SDSB), single-driven multi-beam (SDMB), multi-driven single-beam (MDSB), and multi-driven multi-beam (MDMB) problems. The MDMB is the most general problem type. Once the MDMB problem is solved, we can use the same algorithm to solve all four problems. Therefore, we focus on the MDMB problem in this work.

Let~$\tilde{e}_{nl}$ denote a target EEP of driven port $n$ toward the direction defined by index~$l$. In this work, we focus on solving the following magnitude minimax-fitting (MMF) problem:
\begin{mini}|l|[0]
    {\mr \in \mathbb{C}^{\NP}}
    {\max_{\substack{n=1,\ldots,\ND, \\l=1,\ldots,L}} \big| |\newe_{nl}(\mr)|^2 - |\tilde{e}_{nl}|^2 \big|}
    {\label{eq:MMF}}
    {}
    \addConstraint{|r_m|}{=1,\quad}{m=1,\ldots,\NP}.
\end{mini}
The aim is to find reflection coefficients $\mr$ for the scatterer ports to minimize the largest error between the resulting EEPs' squared magnitudes and target EEPs' squared magnitudes. The function that is minimized is called the cost function.

Utilizing the MMF formulation is a key improvement compared to the approach presented in~\cite{Salmi2024}. Suppose only the sum of squared EEP magnitudes is maximized. In that case, the resulting EEPs potentially have very different magnitudes, which consequently requires feeding coefficient amplitude tapering to achieve realized-gain optimal beam steering~\cite{Salmi2023}. Feeding the elements with varying magnitudes of signal increases losses because the efficiency of amplifiers decreases at low power levels. Therefore, we use the MMF formulation and the target EEP magnitudes are the same for all driven ports~$n$, that is, $|\tilde{e}_{nl}|^2~=~|\tilde{e}_{n'l}|^2~\quad~\forall n,n'~\in~[1,~\ND]$. 

The challenge in multi-beam optimization is that we do not know the phase distribution of the target EEPs. That is, although the target pattern~$\tilde{e}_{nl}$ in~\eqref{eq:MMF} is a complex quantity, only the squared magnitude of it, $|\tilde{e}_{nl}|^2$ is given. The same issue also appears in the antenna array beamforming, where we want to find feeding coefficients of antenna elements to form a shaped array radiation pattern~\cite{Cao2017, Kassakian2005, Boyd1997}. 

The MMF formulation has been used in pattern synthesis in~\cite{Boyd1997} and~\cite{Cao2017}, for instance. Alternative approaches include magnitude least squares fitting~\cite{Kassakian2005} and fitting the magnitude pattern within given boundaries~\cite{Fuchs2014}. The MMF formulation is used in this work because it can be implemented using available optimization tools, and the same formulation can be used in all four problem types. 

\subsection{Local solutions via optimization on Riemannian manifold}
\label{sec:method_mo}
The reflection coefficients are constant modulus constrained, that is, the magnitudes of the reflection coefficients are fixed because the scatterer port terminations are desired to be reactive. Reactive terminations can be implemented using short- or open-ended transmission lines~\cite{Salmi2024_pca}. In addition, we want to avoid resistive terminations to reduce losses. Using the optimization on a Riemannian manifold, a local solution is efficiently found regardless of the constant modulus constraint~\cite{Salmi2024, Zhong2022MIMO, Zhong2022}.

A conventional constrained optimization algorithm, such as the Lagrangian method of multipliers or a barrier method, would penalize the cost function if the variables are far from the feasible region. As the feasible region is non-convex (constant modulus constraints), the conventional methods encounter difficulties staying in the feasible region while searching for local extrema. In addition, much parameter tuning is often required to ensure convergence within a reasonable time frame.

Optimization on a Riemannian manifold (\textit{manifold optimization, MO}) searches for the solution directly from the feasible region, assuming that the region is a smooth Riemannian manifold~\cite{boumal2023_book, absil2008_book}. Similar to many optimization algorithms in Euclidean space, the gradient of the cost function defines the direction which to select the next iteration point. Instead of computing the gradient in Euclidean space, the intrinsic Riemannian gradient is used in the manifold optimization. In other respects, the same algorithms can be used in manifold optimization as in Euclidean optimization.

In our problem setting, the Riemannian manifold is a Cartesian product of $\NP$ complex unit circles:
\begin{align}
    \mathcal{C}^{\NP} = \{ \bm{x} \in \mathbb{C}^{\NP} : \; |x_m| = 1, \forall m=1,\ldots,\NP\}.
\end{align}
The manifold optimization on the manifold~$\mathcal{C}^{\NP}$ has been demonstrated in~\cite{Salmi2024, Zhong2022, Alhujaili2019}, where the key algorithmic steps are also explained.

With the MMF formulation and manifold optimization framework, the problem is
\begin{mini}|l|[0]
    {t \in \mathbb{R}, \mr \in \mathcal{C}^{\NP}}
    {t}
    {\label{eq:manopt_MMF}}
    {}
    \addConstraint{-t \leq |\newe_{nl} (\mr)|^2 - |\tilde{e}_{nl}|^2}{\leq t, \quad}{\forall l,n}.
\end{mini}
We use the Riemannian augmented Lagrangian method (RALM) for solving \eqref{eq:manopt_MMF}. The algorithm is detailed in~\cite{Liu2020}. As the sub-solver in the RALM algorithm, we use the Riemannian Broyden–Fletcher–Goldfarb–Shanno algorithm (RBFGS)~\cite{Huang2016, Diehl2010}. The solver parameters, augmented Lagrangian function, and its gradient are given in Appendix~\hyperref[app:A]{A}.

\subsection{Fundamental bound via semi-definite relaxation}
\label{sec:method_sdr}
While the manifold optimization algorithm does yield a local solution for~\eqref{eq:MMF}, it does not provide information on the proximity of this solution to the global optimum. Consequently, it becomes a challenge to determine whether the local search should be rerun or whether the derived solution is already near-optimal.

With the application of semi-definite relaxation (SDR), we can approximate the original non-convex problem as a convex problem. The unique solution to the convex problem has a lower cost function value than the global solution to the original minimization problem~\cite{Wang2020}. Therefore, the cost function value of the SDR solution is bound to the original problem. The bound is upper for maximization problems and lower for minimization problems. By comparing local solutions with this bound, we can assess the proximity of these local solutions to the global optimum. In addition, with the help of the bound, the potential performance of the designed antenna topology can be evaluated without running the local optimizer. 

We first reformulate~\eqref{eq:MMF} as a quadratically constrained linear program (QCLP), where the unknown~$\bm{x}$ is a vector of incident waves to the scatterer ports. Because the matrices in the quadratic constraint functions are not all positive-definite, the QCLP is not necessarily convex \cite{Boyd2004_cvx}. Therefore we apply SDR to the QCLP introducing a new matrix-variable $\bm{X}$ as the quadratic terms of form $\bm{x}^{\mathrm{H}} \bm{Q} \bm{x}$ are replaced with linear terms $\trace(\bm{Q} \bm{X})$. The constraint $\bm{X} = \bm{x} \bm{x}^{\mathrm{H}}$ is dropped resulting in a convex problem \cite{Luo2010}. The convex relaxed QCLP is
\begin{mini!}|l|[1]
    {t \in \mathbb{R}, \bm{x} \in \mathbb{C}^{J}, \bm{X} \in \mathbb{C}^{J \times J} }
    {t}
    {\label{eq:SDR_MMF}}
    {}
    \addConstraint{-t \leq f_{i}(\bm{x}, \bm{X}) }{\leq t, \quad}{\forall i \in [1, \ND L] \label{eq:SDR_minimaxconst}}
    \addConstraint{ g_{j}(\bm{x}, \bm{X}) }{= 0, \quad }{j \in [1,J] \label{eq:SDR_passivityconst}}
    \addConstraint{ h_{j}(\bm{x}, \bm{X}) }{= 0, \quad }{j \in [1,J] \label{eq:SDR_independenceconst}}
    \addConstraint{ \begin{bmatrix}
        \bm{X} & \bm{x} \\ \bm{x}^{\mathrm{H}} & 1
    \end{bmatrix}}{\succeq 0.}{\label{eq:SDR_posdefconst}}
\end{mini!}
The detailed derivation of the SDR problem, as well as the constraint function definitions $f_i$, $g_j$, and $h_j$, are given in Appendix~\hyperref[app:B]{B}.

The constraint \eqref{eq:SDR_minimaxconst} ensures that the gap between the resulting EEP magnitudes and the target EEP magnitudes is smaller than $t$. The constraint \eqref{eq:SDR_passivityconst} ensures that the scatterer port terminations are reactive. The constraint \eqref{eq:SDR_independenceconst} ensures that the terminations of the scatterer ports do not depend on the driven ports, that is, the same set of terminations is used for all driven ports. The problem is vectorized, and $J=\ND \NP$.

Let~$(t^{\star}, \bm{x}^{\star}, \bm{X}^{\star})$ be the solution to~\eqref{eq:SDR_MMF}. We aspire for the solution of the problem to satisfy the condition \mbox{$\bm{X}^{\star} = \bm{x}^{\star} (\bm{x}^{\star})^{\mathrm{H}}$}, that is, $\bm{X}^{\star}$ being a rank-1 matrix. However, this condition typically does not hold, and a realizable result vector needs to be solved from~$\bm{X}^{\star}$ and~$\bm{x}^{\star}$. The complexity of extracting this solution is contingent on the accuracy of the SDR approximation, which in turn is dependent on the nature of the problem~\cite{Luo2010}. In specific problem contexts, such as the one addressed in~\cite{Corcoles2015}, it is possible to extract a realizable result from the SDR solution successfully. However, for large-scale issues like the MDMB problem, obtaining a realizable result becomes exceedingly difficult due to the diminished approximation accuracy. 

In this work, SDR is employed primarily to establish a boundary for the original problem instead of procuring a realizable result. We visualize this boundary by calculating the EEPs at selective points $l \in [1, L]$ using the function $f_{i}(\bm{x}^{\star}, \bm{X}^{\star})$. For illustrative purposes, we also try to extract the realizable SDR result obtained from the bound solution, as elucidated in Appendix~\hyperref[app:B]{B}.

In the test cases, our SDR implementation is compared to the approach of \cite{Corcoles2015}, which we call the minimum-power SDR (MP-SDR) formulation. The main difference is that we have extended the formulation to cover multiple target directions with fixed reactive loads. In addition, we optimize realized gain instead of directivity. We assume that the generator impedances of the driven ports are fixed, and no matching circuits will be added to the antenna. We, therefore, optimize the matching as well.

\section{Demonstration using connected bowtie-slot antenna}
\label{sec:demonstration}

The optimization framework is demonstrated using the connected bowtie-slot antenna. We analyze all four cases. The SDSB case is demonstrated with two target directions for gain maximization, $\theta_0 = 0^{\circ}$ and $\theta_0 = 20^{\circ}$, and the solution of the $\theta_0 = 20^{\circ}$ case is verified by simulation in which the scatterer ports are terminated to microstrip lines with appropriate lengths. The SDMB case demonstrates a design of an antenna with a flat-top radiation pattern from $\theta = -60^{\circ}$ to $\theta = 60^{\circ}$. The MDSB case maximizes the antenna array's gain toward both $\theta_0 = 0^{\circ}$ and $\theta_0 = 20^{\circ}$, separately. Finally, the MDMB case demonstrates a design of a grating-lobe-free sparse antenna array with $1.5$-$\lambda$ inter-element distances and $\theta \in \left[-19.5^{\circ}, 19.5^{\circ} \right]$ field of view. The realizability of the MDMB solution is tested by a simulation.

In this section, we focus on demonstrating the four types of problems and showing that reactively loaded antennas and arrays can be synthesized using different optimization approaches. We expect all the tested algorithms to perform well in the simplest problems, such as in the SDSB. However, as the complexity increases, manifold optimization is anticipated to dominate in performance. The analysis and interpretation of the results are deferred to Section~\ref{sec:discussion}.

\subsection{Test setup}
\label{sec:demonstration_setup}
\begin{figure}[t]
     \centering
     \subfloat[]{
        \includegraphics[trim=0cm 0cm 0cm 0cm, clip, width=0.55\linewidth]{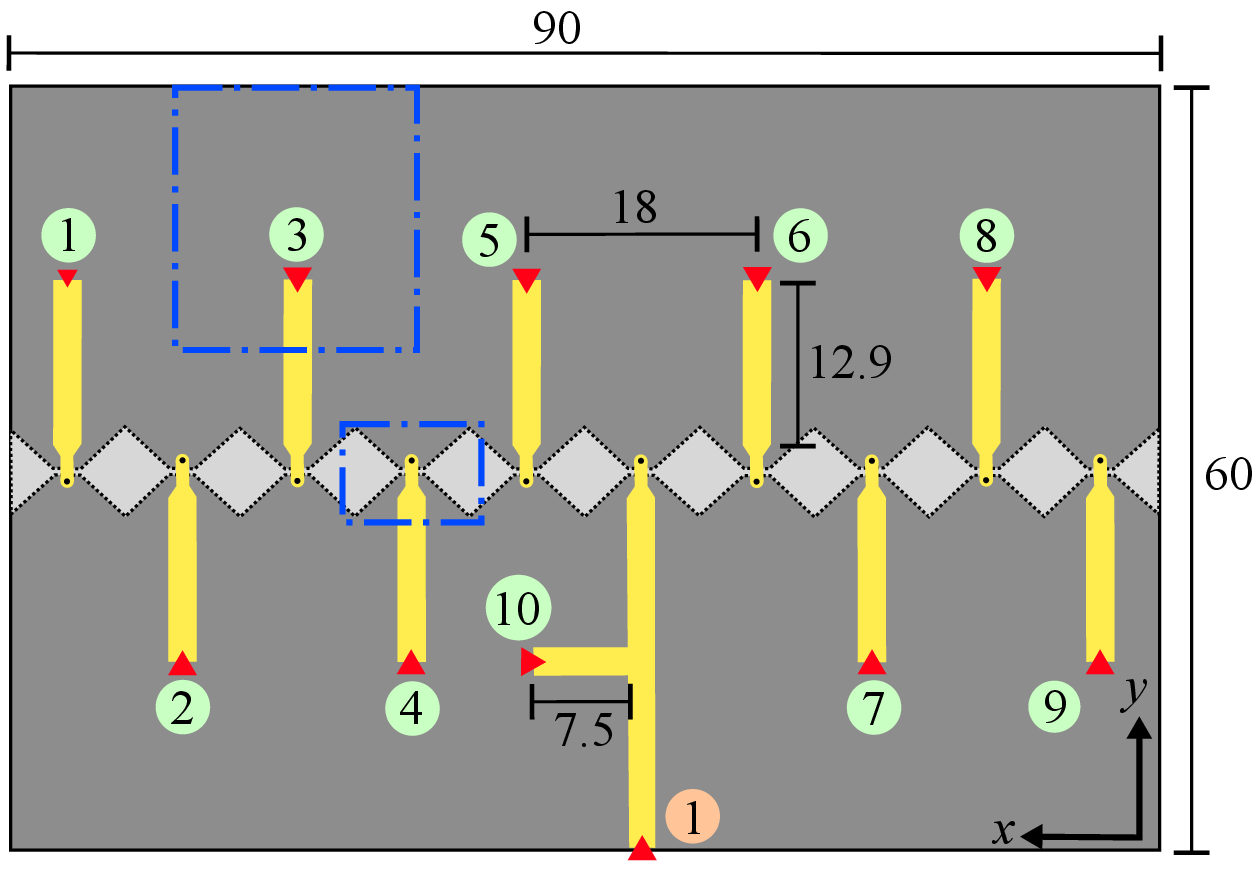}
        \label{fig:element_full}
     }
     \hfill
     \subfloat[]{
        \includegraphics[trim=0cm 0cm 0cm 0cm, clip, width=0.35\linewidth]{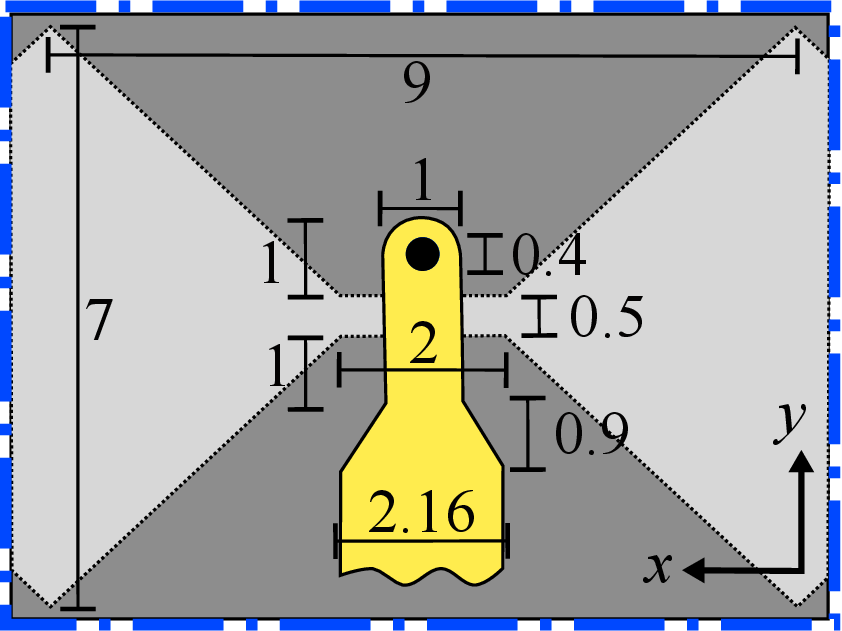}
        \label{fig:element_zoom}
     } \\
     \subfloat[]{
        \includegraphics[trim=0.3cm 0cm 0.2cm 0cm, clip, width=0.8\linewidth]{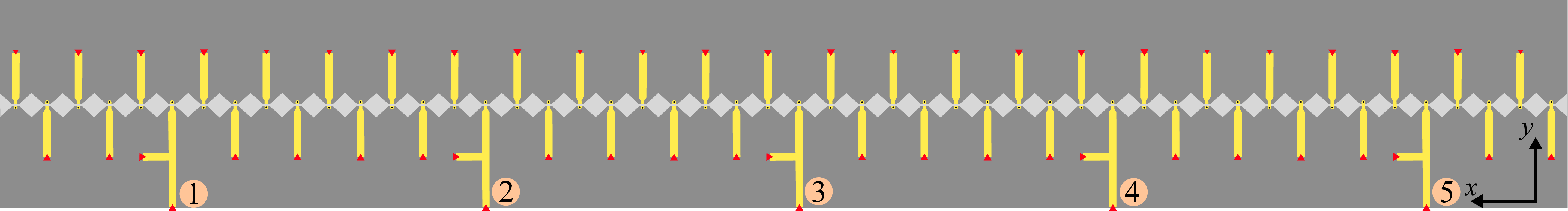}
        \label{fig:array_full}
     } \hfill
     \subfloat[]{
        \includegraphics[trim=0cm 0cm 0cm 0cm, clip, width=0.14\linewidth]{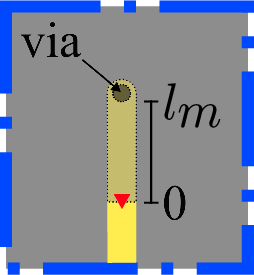}
        \label{fig:feedline_termination}
     }
    \caption{(a) Single-driven antenna. (b) Zoomed view of the feeding structure and slot. (c) Multi-driven antenna array. (d) Termination of the scatterer port illustrated. Figures are from the bottom view. All the dimensions are in millimeters. The overall dimensions for the array in (c) are $450\, \mathrm{mm} \times 60\, \mathrm{mm}$.}
    \label{fig:test_antenna}
\end{figure}

The single-driven test antenna is illustrated in Fig.~\ref{fig:element_full}. The multi-driven antenna array is in Fig.~\ref{fig:array_full}. The figures illustrate the initial simulation model where both driven and scatterer ports are excited. The simulator is CST Microwave Studio \cite{cst}. The simulation models of the presented antennas are provided as supplementary material.

In these demonstrations, the multi-driven antenna is constructed by repeating the base design of the single-driven antenna five times along the $x$-axis. However, the optimized terminations of the scatterer ports are generally non-periodic. Even the base design could be made entirely non-periodic without any modification of the proposed optimization scheme.

The antenna is designed by following the guidelines for the reactively loaded antenna arrays presented in \cite{Salmi2024}. It consists of two layers of copper and a substrate between the copper layers. Figure~\ref{fig:test_antenna} illustrates the bottom copper layer in yellow, and the top copper in dark grey. The light grey color illustrates the slot cut in the top copper. The antenna does not have a separate ground plane, but the top copper represents the ground for the microstrip lines.

The substrate is 1.52-mm-thick Rogers Ro4360G2 with a dielectric constant of 6.4 and a dissipation factor of 0.0038. The designed operation frequency is 5\,GHz. The vias between the top and bottom copper layers are marked as black dots in Fig.~\ref{fig:test_antenna}. The key dimensions of the antennas are shown in Fig.~\ref{fig:element_zoom}.

The ports, illustrated as red triangles in Fig.~\ref{fig:test_antenna}, are aligned between the microstrip line ends and the top copper. The single-driven antenna has 10~scatterer ports implemented using 50-$\Omega$ microstrip lines. The numbering of the scatterer ports is marked using green circles, and the driven port is highlighted in orange. In the multi-driven antenna in Fig.~\ref{fig:array_full}, the driven ports are numbered from 1 to 5, ascending toward the negative $x$-axis.

The reflection coefficients are realized after the optimization by tuning the lengths of the microstrip lines and shorting them, as illustrated in Fig.~\ref{fig:feedline_termination}. The lengths of the short-circuited transmission lines, $l_m$, are calculated based on the optimized reflection coefficients~$\rc_m$ as
\begin{align}
    l_m = \frac{1}{\beta} \tan^{-1} \bigg( \frac{z_0}{\ji z_{\mathrm{line}} } \frac{1 + \rc_m}{1 - \rc_m} \bigg),
\end{align}
where $\beta = 230\, \mathrm{rad}/\mathrm{m}$ is the phase constant of the used microstrip line. The reference impedance, which is also the generator impedance and used in \eqref{eq:r_vs_z}, is $z_0 = 50\, \Omega$, and $z_{\mathrm{line}} = 48\, \Omega$ is the wave impedance of the microstrip line. The length $l_m$ being negative indicates that the line should be shortened from its initial length. 

The transmission lines are assumed to be lossless, corresponding to reactive loads in the optimization. In practice, the losses in the scatterer port terminations are inevitable, but we assume that the delays caused by the terminations have much stronger effects than their attenuation. The full-wave electromagnetic simulations show the insignificant effect of the losses.

In the following four test cases, the fundamental bound is computed using SDR as described in Section~\ref{sec:method_sdr}, and the realizable solution is obtained with the manifold optimization algorithm described in Section~\ref{sec:method_mo}. We utilize the CVX toolbox with SeDuMi solver~\cite{cvx, sedumi} for solving the semi-definite program, and the Manopt toolbox~\cite{manopt} for the manifold optimization. The optimization codes are provided as supplementary material.

As a reference, the problems are also solved using a genetic algorithm (GA) implemented in the MATLAB library~\cite{matlab}.  In the single-beam cases, we further compare our optimization methods to the MP-SDR approach presented in~\cite{Corcoles2015}.

The local search algorithms, MO and GA, are run 10 times for every optimization problem, starting with different initial guesses for $\mr$ in each run. The initial guess for the first run is derived from the fundamental bound, as described in Appendix~\hyperref[app:B]{B}. In single-beam problems, the second initial guess is obtained from the MP-SDR optimization solution. The remaining initial guesses are selected randomly. This process generates multiple local solutions, from which we analyze the worst, average, and best outcomes.

The target EEP magnitudes, $|\tilde{e}_{nl}|^2$, are chosen ad hoc based on empirical testing and practical results, with the only requirement being that they exceed achievable EEP magnitudes. The co-polarization of EEPs is chosen based on Ludwig's third definition \cite{Ludwig1973}.

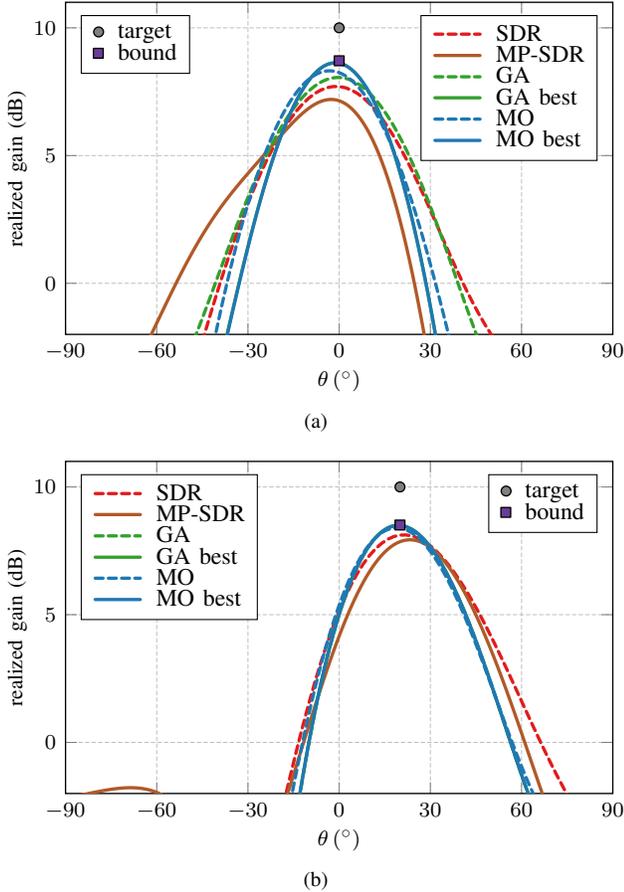
\begin{figure}[t]
     \centering
     \subfloat[]{
        \begin{tikzpicture}

\begin{axis}[%
width=\columnwidth,
height=6cm,
at={(0cm,0cm)},
xmin=-90,
xmax=90,
xlabel={$\theta \, (^{\circ})$},
ymin=-2,
ymax=11,
ylabel={realized gain (dB)},
ylabel shift = -0.15cm,
xlabel shift = -0.1cm,
xtick = {-90, -60, ..., 90},
xmajorgrids,
ymajorgrids,
grid style={dashed, cap=round, dash pattern = on 0.5mm off 0.375mm, line width=1/4, opacity=0.5},
axis background/.style={fill=white},
axis line style={solid, line width=0.5pt},
]

\addplot[target, only marks] table[]{Figures/single-driven-single-beam/SDSB_result_data/SDSB_result_scan0-8.tsv};
\label{SDSB_scan0:target}


\addplot[SDR, densely dashed] table[]{Figures/single-driven-single-beam/SDSB_result_data/SDSB_result_scan0-2.tsv};
\label{SDSB_scan0:SDR}

\addplot[MPSDR] table[]{Figures/single-driven-single-beam/SDSB_result_data/SDSB_result_scan0-3.tsv};
\label{SDSB_scan0:MPSDR}

\addplot[GA, densely dashed] table[]{Figures/single-driven-single-beam/SDSB_result_data/SDSB_result_scan0-5.tsv};
\label{SDSB_scan0:GA}

\addplot[GA] table[]{Figures/single-driven-single-beam/SDSB_result_data/SDSB_result_scan0-4.tsv};
\label{SDSB_scan0:GAbest}

\addplot[ManuOpt, densely dashed] table[]{Figures/single-driven-single-beam/SDSB_result_data/SDSB_result_scan0-7.tsv};
\label{SDSB_scan0:MO}

\addplot[ManuOpt] table[]{Figures/single-driven-single-beam/SDSB_result_data/SDSB_result_scan0-6.tsv};
\label{SDSB_scan0:MObest}

\addplot[bound, only marks] table[]{Figures/single-driven-single-beam/SDSB_result_data/SDSB_result_scan0-1.tsv};
\label{SDSB_scan0:bound}

\begin{pgfonlayer}{foreground layer}
	\node[draw,fill=white,draw=black,cap=round, fill opacity=0.95] at (axis cs: -85, 10.5) [anchor=north west] {\begin{tabular}{ccl}
			\ref{SDSB_scan0:target} & \quad & \small{target} \\
			\ref{SDSB_scan0:bound} & & \small{bound}
	\end{tabular}};
\end{pgfonlayer}

\begin{pgfonlayer}{foreground layer}
	\node[draw,fill=white,draw=black,cap=round, fill opacity=0.95] at (axis cs: 85, 10.5) [anchor=north east] {\begin{tabular}{ccl}
			\ref{SDSB_scan0:SDR} & \quad & \small{SDR} \\
            \ref{SDSB_scan0:MPSDR} & & \small{MP-SDR} \\
			\ref{SDSB_scan0:GA} & & \small{GA} \\
			\ref{SDSB_scan0:GAbest} & & \small{GA best} \\
			\ref{SDSB_scan0:MO} & & \small{MO} \\
			\ref{SDSB_scan0:MObest} & & \small{MO best}
	\end{tabular}};
\end{pgfonlayer}

\end{axis}
\end{tikzpicture}%
        \label{fig:SDSB_result_scan0}
     } \\
     \subfloat[]{
    \begin{tikzpicture}

\begin{axis}[%
width=\columnwidth,
height=6cm,
at={(0cm,0cm)},
xmin=-90,
xmax=90,
xlabel={$\theta \, (^{\circ})$},
ymin=-2,
ymax=11,
ylabel={realized gain (dB)},
ylabel shift = -0.15cm,
xlabel shift = -0.1cm,
xtick = {-90, -60, ..., 90},
xmajorgrids,
ymajorgrids,
grid style={dashed, cap=round, dash pattern = on 0.5mm off 0.375mm, line width=1/4, opacity=0.5},
axis background/.style={fill=white},
axis line style={solid, line width=0.5pt},
]

\addplot[target, only marks] table[]{Figures/single-driven-single-beam/SDSB_result_data/SDSB_result_scan20-8.tsv};
\label{SDSB_scan20:target}


\addplot[SDR, densely dashed] table[]{Figures/single-driven-single-beam/SDSB_result_data/SDSB_result_scan20-2.tsv};
\label{SDSB_scan20:SDR}

\addplot[MPSDR] table[]{Figures/single-driven-single-beam/SDSB_result_data/SDSB_result_scan20-3.tsv};
\label{SDSB_scan20:MPSDR}

\addplot[GA, densely dashed] table[]{Figures/single-driven-single-beam/SDSB_result_data/SDSB_result_scan20-5.tsv};
\label{SDSB_scan20:GA}

\addplot[GA] table[]{Figures/single-driven-single-beam/SDSB_result_data/SDSB_result_scan20-4.tsv};
\label{SDSB_scan20:GAbest}

\addplot[ManuOpt, densely dashed] table[]{Figures/single-driven-single-beam/SDSB_result_data/SDSB_result_scan20-7.tsv};
\label{SDSB_scan20:MO}

\addplot[ManuOpt] table[]{Figures/single-driven-single-beam/SDSB_result_data/SDSB_result_scan20-6.tsv};
\label{SDSB_scan20:MObest}

\addplot[bound, only marks] table[]{Figures/single-driven-single-beam/SDSB_result_data/SDSB_result_scan20-1.tsv};
\label{SDSB_scan20:bound}

\begin{pgfonlayer}{foreground layer}
	\node[draw,fill=white,draw=black,cap=round, fill opacity=0.95] at (axis cs: 85, 10.5) [anchor=north east] {\begin{tabular}{ccl}
			\ref{SDSB_scan20:target} & \quad & \small{target} \\
			\ref{SDSB_scan20:bound} & & \small{bound}
	\end{tabular}};
\end{pgfonlayer}

\begin{pgfonlayer}{foreground layer}
	\node[draw,fill=white,draw=black,cap=round, fill opacity=0.95] at (axis cs: -85, 10.5) [anchor=north west] {\begin{tabular}{ccl}
			\ref{SDSB_scan20:SDR} & \quad & \small{SDR} \\
            \ref{SDSB_scan20:MPSDR} & & \small{MP-SDR} \\
			\ref{SDSB_scan20:GA} & & \small{GA} \\
			\ref{SDSB_scan20:GAbest} & & \small{GA best} \\
			\ref{SDSB_scan20:MO} & & \small{MO} \\
			\ref{SDSB_scan20:MObest} & & \small{MO best}
	\end{tabular}};
\end{pgfonlayer}

\end{axis}
\end{tikzpicture}%
    \label{fig:SDSB_result_scan20}
     }
    \caption{Realized gain patterns of the single-driven antenna on $\varphi = 0^{\circ}$ plane optimized using different methods to maximize the gain towards the directions (a) $\theta = 0^{\circ}$, (b) $\theta = 20^{\circ}$.}
    \label{fig:SDSB_result}
\end{figure}


\subsection{Study 1: Single-driven single-beam}
\label{sec:demonstration_sdsb}
The SDSB problem appears when designing reactively controlled antenna arrays and ESPARs. Furthermore, it arises in the design of reconfigurable intelligent surfaces for single-user configurations \cite{Wu2020}. Within these cases, driven elements are perceived as incident plane waves originating from diverse angles, yet the formulation generally parallels that used in antenna arrays.

\begin{figure}[t]
    \includegraphics[width=\linewidth]{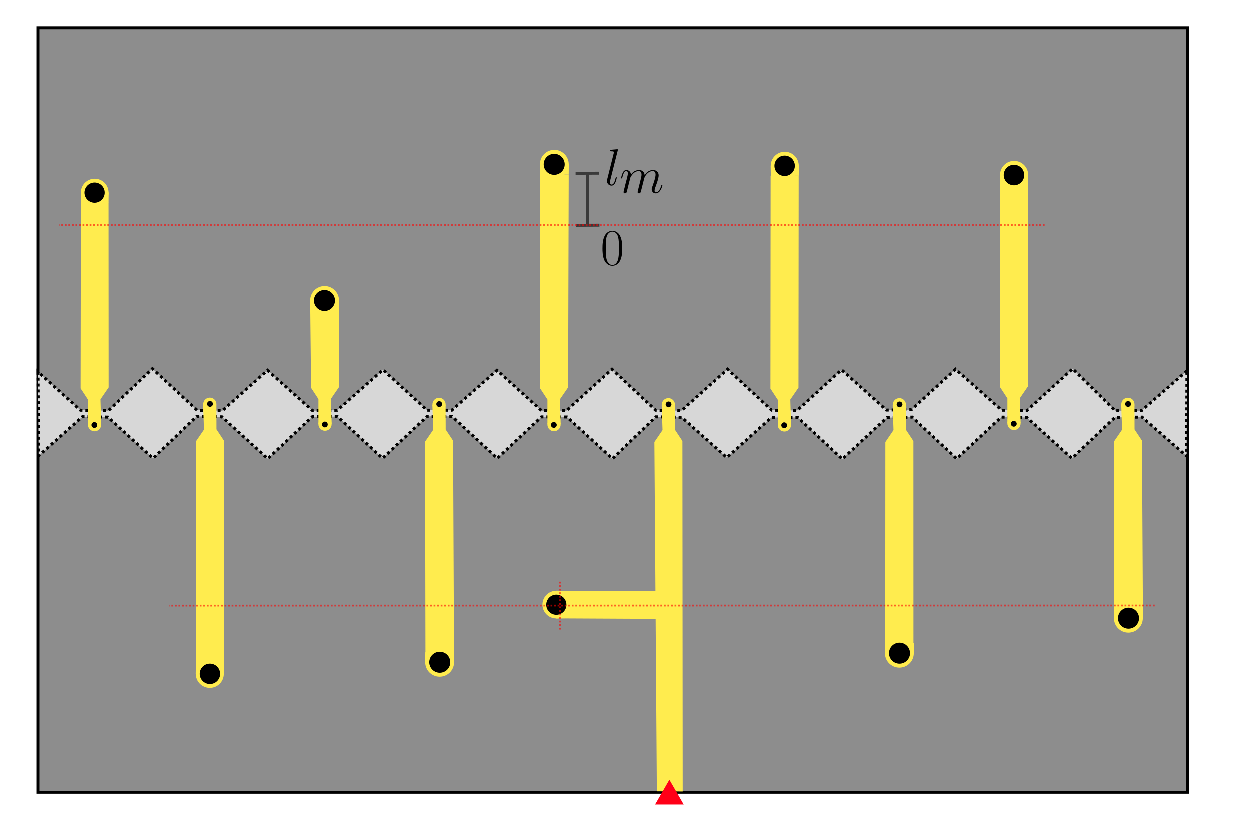}
    \caption{Simulation model of the antenna in which the scatterer ports are terminated to optimal length shorted microstrip lines. }
    \label{fig:SDSB_result_antenna}
\end{figure}

We optimize the lengths of the scatterer port termination lines concerning two distinct targets. The directions to which the gain is maximized are $\theta_0 = 0^{\circ}$, and $\theta_0 = 20^{\circ}$. 

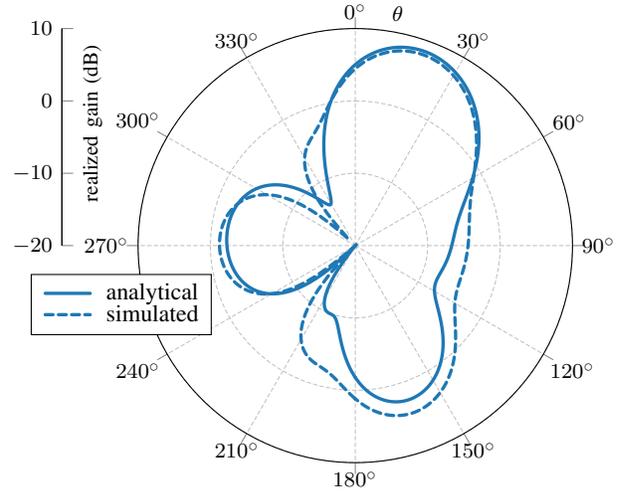
\begin{figure}[t]
    \begin{tikzpicture}

\begin{polaraxis}[
    width=\linewidth-1.5cm,
    height=\linewidth-1.5cm,
    xticklabel=$\pgfmathprintnumber{\tick}^\circ$,
    xlabel=$\theta$,
    xlabel style = {at={(axis cs:15, 11)}, rotate=45},
    ylabel = realized gain (dB),
    ylabel style = {at={(axis cs:295, 20)}, anchor=center, rotate=90},
    xtick={0,30,...,330},
    ytick={-20, -10, 0, 10},
    ymin=-20, ymax=10,
    y coord trafo/.code=\pgfmathparse{#1+20},
    rotate=-90,
    y coord inv trafo/.code=\pgfmathparse{#1-20},
    x dir=reverse,
    xticklabel style={anchor=-\tick-90},
    yticklabel style={anchor=east, xshift=-3.9cm},
    y axis line style={yshift=-3.9cm},
    ytick style={yshift=-3.9cm},
    grid style={dashed, cap=round, dash pattern = on 0.5mm off 0.375mm, line width=1/4, opacity=0.5},
    legend style={at={(0.72,0)}, anchor=south}
    ]

\addplot[data cs = polarrad, ManuOpt] table[col sep=comma]{Figures/single-driven-single-beam/SDSB_result_data_simu/SDSB_polardata_analytical.txt};
\label{SDSB_simu:analytical}

\addplot[data cs = polarrad, ManuOpt, densely dashed] table[col sep=comma]{Figures/single-driven-single-beam/SDSB_result_data_simu/SDSB_polardata_simu.txt};
\label{SDSB_simu:simu}

\begin{pgfonlayer}{foreground layer}
	\node[draw,fill=white,draw=black,cap=round, fill opacity=0.95] at (axis cs: 265, 25) [anchor=north west] {\begin{tabular}{ccl}
			\ref{SDSB_simu:analytical} & \quad & \small{analytical} \\
			\ref{SDSB_simu:simu} & & \small{simulated}
	\end{tabular}};
\end{pgfonlayer}
        
\end{polaraxis}

\end{tikzpicture}%
    \caption{Resulting realized gain pattern of the single-driven antenna when the radiation is optimized to $\theta = 20^{\circ}$. Comparison to simulated result. }
    \label{fig:SDSB_simu}
\end{figure}

\begin{table}[t]
    \centering
    \caption{Optimal reflection coefficients ($r_m$) and corresponding shorted microstrip line lengths ($l_m$) of the SDSB $\theta_0 = 20^{\circ}$ problem.}
    \begin{tabular}{ccccccccccc}
        $m$ & $1$ & $2$ & $3$ & $4$ & $5$ & $6$ & $7$ & $8$ & $9$ & $10$ \\ \toprule
        $\angle r_m \,(^{\circ})$ & $136$ & $63$ & $-3$ & $89$ & $81$ & $84$ & $107$ & $104$ & $175$ & $-166$ \\[0.3em]
        $l_m$ (mm) & $1.7$ &$ 4.5$ & $-6.7$ & $3.5$ & $3.8$ & $3.7$ & $2.9$ & $3.0$ & $0.2$ & $-0.5$ \\ \bottomrule
    \end{tabular}
    \label{tab:SDSB_r_and_tl}
\end{table}

Figure \ref{fig:SDSB_result} illustrates the antenna's realized gain pattern in $\varphi = 0^{\circ}$ plane after terminating the scatterer ports. The results are computed analytically with MATLAB. In both test cases, the target is 10\,dB realized gain toward the desired direction. The dashed lines of GA and MO indicate the expected results, that is, the results that have median cost function values amongst the optimization runs. The solid MO and GA results indicate the best-obtained results.

The MO gives the best results in both cases. Also, the GA converges to the same solutions as the MO. The expected results of MO and GA are very close to the best results. 

The SDR result refers to the realizable SDR solution which is extracted from the bound. The MP-SDR result is computed using the method presented in \cite{Corcoles2015}. Both the realizable SDR solution and the MP-SDR solution underperform compared to the bound and local solvers' solutions.

The best MO result in case $\theta_0 = 20^{\circ}$ is confirmed by antenna simulations. Fig.~\ref{fig:SDSB_result_antenna} shows the simulated antenna, where the scatterer ports are terminated to microstrip lines with lengths corresponding to the optimal reflection coefficients. Table \ref{tab:SDSB_r_and_tl} lists the reflection coefficient angles and microstrip line lengths. The shorting vias have a radius of 0.8\,mm. The driven port is illustrated as a red triangle, and the initial locations of the scatterer port excitations are highlighted as red dashed lines.

Figure \ref{fig:SDSB_simu} compares the simulator-validated and analytically computed realized gain patterns of the optimized single-driven antenna. The analytical patterns are computed with~\eqref{eq:new_enl}. The patterns align well in the target direction but show slight deviations elsewhere. This discrepancy arises from losses in transmission lines and the other effects of microstrip lines, such as surface waves. At low gain levels, the effects of these approximations become visible.


\subsection{Study 2: Single-driven multi-beam}
\label{sec:demonstration_sdmb}
The SDMB problems appear in antennas when the scatterer port loads cannot be tuned. A typical application is the synthesis of a single-fed antenna with the desired radiation pattern. In addition, the SDMB appears in antenna array synthesis if the array is large and can be modeled using periodic boundary conditions, that is, unit-cell simulations. In this case, the designed antenna rather exhibits a shaped beam than a multi-beam situation because unit-cell simulation is not used.

\begin{figure}[t]
    \begin{tikzpicture}

\begin{axis}[%
width=\columnwidth,
height=6cm,
at={(0cm,0cm)},
xmin=-90,
xmax=90,
xlabel={$\theta \, (^{\circ})$},
ymin=-3,
ymax=7,
ylabel={realized gain (dB)},
ylabel shift = -0.15cm,
xlabel shift = -0.1cm,
xtick = {-90, -60, ..., 90},
xmajorgrids,
ymajorgrids,
grid style={dashed, cap=round, dash pattern = on 0.5mm off 0.375mm, line width=1/4, opacity=0.5},
axis background/.style={fill=white},
axis line style={solid, line width=0.5pt},
]

\addplot[target] table[]{Figures/single-driven-multi-beam/SDMB_result_data/SDMB_result-7.tsv};
\label{SDMB:target}


\addplot[SDR, densely dashed] table[]{Figures/single-driven-multi-beam/SDMB_result_data/SDMB_result-2.tsv};
\label{SDMB:SDR}

\addplot[GA, densely dashed] table[]{Figures/single-driven-multi-beam/SDMB_result_data/SDMB_result-3.tsv};
\label{SDMB:GA}

\addplot[GA] table[]{Figures/single-driven-multi-beam/SDMB_result_data/SDMB_result-4.tsv};
\label{SDMB:GAbest}

\addplot[ManuOpt, densely dashed] table[]{Figures/single-driven-multi-beam/SDMB_result_data/SDMB_result-5.tsv};
\label{SDMB:MO}

\addplot[ManuOpt] table[]{Figures/single-driven-multi-beam/SDMB_result_data/SDMB_result-6.tsv};
\label{SDMB:MObest}

\addplot[bound, densely dotted] table[]{Figures/single-driven-multi-beam/SDMB_result_data/SDMB_result-1.tsv};
\label{SDMB:bound}

\begin{pgfonlayer}{foreground layer}
	\node[draw,fill=white,draw=black,cap=round, fill opacity=0.95] at (axis cs: -10, -2.5) [anchor=south east] {\begin{tabular}{ccl}
			\ref{SDMB:target} & \quad & \small{target} \\
			\ref{SDMB:bound} & & \small{bound}
	\end{tabular}};
\end{pgfonlayer}

\begin{pgfonlayer}{foreground layer}
	\node[draw,fill=white,draw=black,cap=round, fill opacity=0.95] at (axis cs: -5, -2.5) [anchor=south west] {\begin{tabular}{ccl}
			\ref{SDMB:SDR} & \quad & \small{SDR} \\
			\ref{SDMB:GA} & & \small{GA} \\
			\ref{SDMB:GAbest} & & \small{GA best} \\
			\ref{SDMB:MO} & & \small{MO} \\
			\ref{SDMB:MObest} & & \small{MO best}
	\end{tabular}};
\end{pgfonlayer}

\end{axis}
\end{tikzpicture}%
    \caption{Realized gain patterns of the single-driven antenna optimized using different methods to maximize the minimum gain towards the desired sector $\theta \in [-60^{\circ}, 60^{\circ}]$. }
    \label{fig:SDMB_result}
\end{figure}
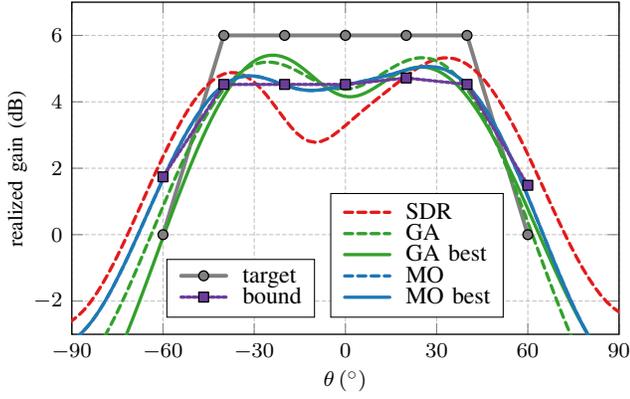

Figure~\ref{fig:SDMB_result} shows the resulting radiation patterns and the target, which is a flat-topped pattern at $\theta \in \left[ -60^{\circ}, 60^{\circ} \right]$ with 6-dB suppression at the edges. The expected MO result overlaps with the best MO result. They are the best solutions for the problem. Also, the GA results in a sufficient EEP, but the realizable SDR is non-satisfactory. The MP-SDR approach is not studied here as the formulation is unsuitable for shaped-beam synthesis.


\subsection{Study 3: Multi-driven single-beam}
\label{sec:demonstration_mdsb}
The MDSB problem appears in the design of a phased antenna array with a reduced number of active excitations. That is, some of the driven elements are replaced by reactively terminated elements whose terminations are tuned when steering the beam.

\begin{figure}[t]
     \centering
     \subfloat[]{
        \begin{tikzpicture}

\begin{axis}[%
width=\columnwidth,
height=6cm,
at={(0cm,0cm)},
xmin=-90,
xmax=90,
xlabel={$\theta \, (^{\circ})$},
ymin=-10,
ymax=21,
ylabel={realized gain (dB)},
ylabel shift = -0.2cm,
xlabel shift = -0.1cm,
xtick = {-90, -60, ..., 90},
xmajorgrids,
ymajorgrids,
grid style={dashed, cap=round, dash pattern = on 0.5mm off 0.375mm, line width=1/4, opacity=0.5},
axis background/.style={fill=white},
axis line style={solid, line width=0.5pt},
]

\addplot[target, only marks] table[]{Figures/multi-driven-single-beam/MDSB_result_data/MDSB_result_scan0-8.tsv};
\label{MDSB_scan0:target}


\addplot[SDR, densely dashed] table[]{Figures/multi-driven-single-beam/MDSB_result_data/MDSB_result_scan0-2.tsv};
\label{MDSB_scan0:SDR}

\addplot[MPSDR] table[]{Figures/multi-driven-single-beam/MDSB_result_data/MDSB_result_scan0-3.tsv};
\label{MDSB_scan0:MPSDR}

\addplot[GA, densely dashed] table[]{Figures/multi-driven-single-beam/MDSB_result_data/MDSB_result_scan0-5.tsv};
\label{MDSB_scan0:GA}

\addplot[GA] table[]{Figures/multi-driven-single-beam/MDSB_result_data/MDSB_result_scan0-4.tsv};
\label{MDSB_scan0:GAbest}

\addplot[ManuOpt, densely dashed] table[]{Figures/multi-driven-single-beam/MDSB_result_data/MDSB_result_scan0-7.tsv};
\label{MDSB_scan0:MO}

\addplot[ManuOpt] table[]{Figures/multi-driven-single-beam/MDSB_result_data/MDSB_result_scan0-6.tsv};
\label{MDSB_scan0:MObest}

\addplot[bound, only marks] table[]{Figures/multi-driven-single-beam/MDSB_result_data/MDSB_result_scan0-1.tsv};
\label{MDSB_scan0:bound}

\begin{pgfonlayer}{foreground layer}
	\node[draw,fill=white,draw=black,cap=round, fill opacity=0.95] at (axis cs: -85, 20) [anchor=north west] {\begin{tabular}{ccl}
			\ref{MDSB_scan0:target} & \quad & \small{target} \\
			\ref{MDSB_scan0:bound} & & \small{bound}
	\end{tabular}};
\end{pgfonlayer}

\begin{pgfonlayer}{foreground layer}
	\node[draw,fill=white,draw=black,cap=round, fill opacity=0.95] at (axis cs: 85, 20) [anchor=north east] {\begin{tabular}{ccl}
			\ref{MDSB_scan0:SDR} & \quad & \small{SDR} \\
            \ref{MDSB_scan0:MPSDR} & & \small{MP-SDR} \\
			\ref{MDSB_scan0:GA} & & \small{GA} \\
			\ref{MDSB_scan0:GAbest} & & \small{GA best} \\
			\ref{MDSB_scan0:MO} & & \small{MO} \\
			\ref{MDSB_scan0:MObest} & & \small{MO best}
	\end{tabular}};
\end{pgfonlayer}

\end{axis}
\end{tikzpicture}%
        \label{fig:MDSB_result_scan0}
     } \\
     \subfloat[]{
        \begin{tikzpicture}

\begin{axis}[%
width=\columnwidth,
height=6cm,
at={(0cm,0cm)},
xmin=-90,
xmax=90,
xlabel={$\theta \, (^{\circ})$},
ymin=-10,
ymax=21,
ylabel={realized gain (dB)},
ylabel shift = -0.2cm,
xlabel shift = -0.1cm,
xtick = {-90, -60, ..., 90},
xmajorgrids,
ymajorgrids,
grid style={dashed, cap=round, dash pattern = on 0.5mm off 0.375mm, line width=1/4, opacity=0.5},
axis background/.style={fill=white},
axis line style={solid, line width=0.5pt},
]

\addplot[target, only marks] table[]{Figures/multi-driven-single-beam/MDSB_result_data/MDSB_result_scan20-8.tsv};
\label{MDSB_scan20:target}


\addplot[SDR, densely dashed] table[]{Figures/multi-driven-single-beam/MDSB_result_data/MDSB_result_scan20-2.tsv};
\label{MDSB_scan20:SDR}

\addplot[MPSDR] table[]{Figures/multi-driven-single-beam/MDSB_result_data/MDSB_result_scan20-3.tsv};
\label{MDSB_scan20:MPSDR}

\addplot[GA, densely dashed] table[]{Figures/multi-driven-single-beam/MDSB_result_data/MDSB_result_scan20-5.tsv};
\label{MDSB_scan20:GA}

\addplot[GA] table[]{Figures/multi-driven-single-beam/MDSB_result_data/MDSB_result_scan20-4.tsv};
\label{MDSB_scan20:GAbest}

\addplot[ManuOpt, densely dashed] table[]{Figures/multi-driven-single-beam/MDSB_result_data/MDSB_result_scan20-7.tsv};
\label{MDSB_scan20:MO}

\addplot[ManuOpt] table[]{Figures/multi-driven-single-beam/MDSB_result_data/MDSB_result_scan20-6.tsv};
\label{MDSB_scan20:MObest}

\addplot[bound, only marks] table[]{Figures/multi-driven-single-beam/MDSB_result_data/MDSB_result_scan20-1.tsv};
\label{MDSB_scan20:bound}

\begin{pgfonlayer}{foreground layer}
	\node[draw,fill=white,draw=black,cap=round, fill opacity=0.95] at (axis cs: 85, 20) [anchor=north east] {\begin{tabular}{ccl}
			\ref{MDSB_scan20:target} & \quad & \small{target} \\
			\ref{MDSB_scan20:bound} & & \small{bound}
	\end{tabular}};
\end{pgfonlayer}

\begin{pgfonlayer}{foreground layer}
	\node[draw,fill=white,draw=black,cap=round, fill opacity=0.95] at (axis cs: -85, 20) [anchor=north west] {\begin{tabular}{ccl}
			\ref{MDSB_scan20:SDR} & \quad & \small{SDR} \\
            \ref{MDSB_scan20:MPSDR} & & \small{MP-SDR} \\
			\ref{MDSB_scan20:GA} & & \small{GA} \\
			\ref{MDSB_scan20:GAbest} & & \small{GA best} \\
			\ref{MDSB_scan20:MO} & & \small{MO} \\
			\ref{MDSB_scan20:MObest} & & \small{MO best}
	\end{tabular}};
\end{pgfonlayer}

\end{axis}
\end{tikzpicture}%
        \label{fig:MDSB_result_scan20}
     }
    \caption{Realized gain patterns of the multi-driven antenna optimized using different methods to maximize the gain towards the directions (a) $\theta = 0^{\circ}$, (b) $\theta = 20^{\circ}$. The driven elements are fed to maximize the realized gain toward the target direction.}
    \label{fig:MDSB_result}
\end{figure}
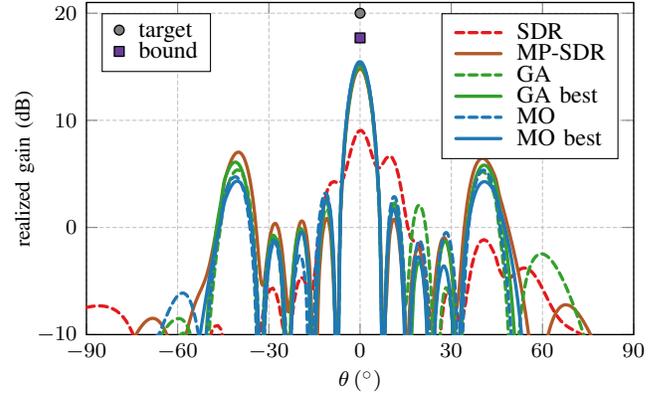
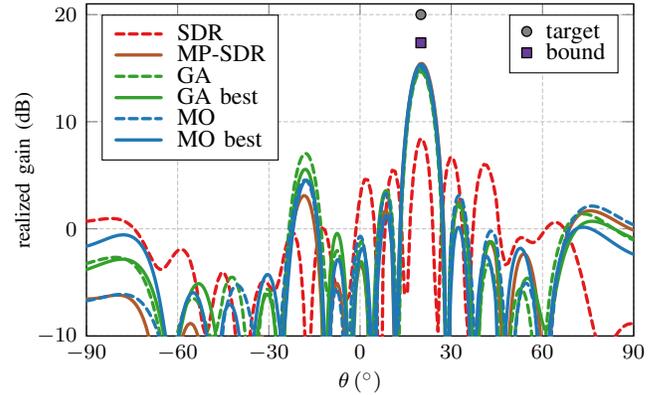

We study the antenna array with five driven ports and 50 scatterer ports shown in Fig.~\ref{fig:array_full}. The first target is to maximize the realized gain of the array toward direction $\theta_0 = 0^{\circ}$, and then toward $\theta_0 = 20^{\circ}$. We maximize the minimum EEP magnitude toward the desired direction and compute the realized gain-optimal feeding coefficients of the driven elements afterward. Consequently, the array gain is maximized without having a large deviation in EEP magnitudes, and the array could be fed using only phase tuning in the elements without amplitude tapering.

The array realized gain patterns, in the case that the five ports are driven optimally, are shown in Fig.~\ref{fig:MDSB_result}. The result curves overlap significantly. The only non-satisfactory result is obtained when trying to realize the SDR-induced fundamental bound. The MP-SDR is more accurate, and the solution can be implemented in practice, resulting in excellent results. The target is visualized as a sum of EEP targets in order to compare the array gain patterns against the target. Similarly, the bound is visualized as a sum of EEP bounds.


\subsection{Study 4: Multi-driven multi-beam}
\label{sec:demonstration_mdmb}
The most complex problem, MDMB, is used in the design of antenna arrays whose elements are not necessarily identical, and unit-cell simulations cannot be utilized. The reactively loaded elements are fixed, that is, they are not tuned when steering the beam. A typical use case is the design of a sparse array. There, the EEPs are shaped to cover a specific sector toward which the beam can be steered without suffering from grating lobes \cite{Salmi2024, skobelev2011book, Maximidis2020}. In addition to antenna array design, multi-beam problems appear in static reflecting surfaces and multi-user reconfigurable intelligent surfaces design \cite{Tirkkonen2024, Wu2020}.

The target EEPs of the MDMB problem are chosen based on the appearance of the grating lobes. We choose the beam-steering area being $\theta_0 \in \left[ -19.5^{\circ}, 19.5^{\circ} \right]$. If the EEPs were flat-topped on that sector and minimal elsewhere, the array would not radiate grating lobes when steering the beam inside that sector \cite{Salmi2024}. 

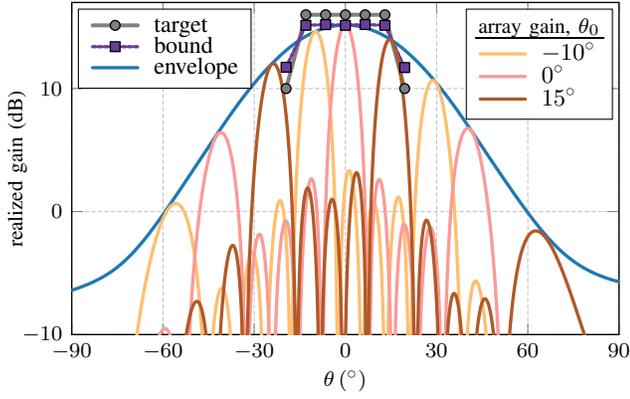
\begin{figure}[t]
    \begin{tikzpicture}

\begin{axis}[%
width=\columnwidth,
height=6cm,
at={(0cm,0cm)},
xmin=-90,
xmax=90,
xlabel={$\theta \, (^{\circ})$},
ymin=-10,
ymax=17,
ylabel={realized gain (dB)},
ylabel shift = -0.3cm,
xlabel shift = -0.1cm,
xtick = {-90, -60, ..., 90},
xmajorgrids,
ymajorgrids,
grid style={dashed, cap=round, dash pattern = on 0.5mm off 0.375mm, line width=1/4, opacity=0.5},
axis background/.style={fill=white},
axis line style={solid, line width=0.5pt},
]

\addplot[target] table[]{Figures/multi-driven-multi-beam/MDMB_result_scan_data/MDMB_result_scan-6.tsv};
\label{MDMB_scan:target}

\addplot[ManuOpt] table[]{Figures/multi-driven-multi-beam/MDMB_result_scan_data/MDMB_result_scan-2.tsv};
\label{MDMB_scan:MObest}

\addplot[bound, densely dotted] table[]{Figures/multi-driven-multi-beam/MDMB_result_scan_data/MDMB_result_scan-1.tsv};
\label{MDMB_scan:bound}

\addplot[resultEEP1] table[]{Figures/multi-driven-multi-beam/MDMB_result_scan_data/MDMB_result_scan-3.tsv};
\label{MDMB_scan:scan1}

\addplot[resultEEP2] table[]{Figures/multi-driven-multi-beam/MDMB_result_scan_data/MDMB_result_scan-4.tsv};
\label{MDMB_scan:scan2}

\addplot[resultEEP3] table[]{Figures/multi-driven-multi-beam/MDMB_result_scan_data/MDMB_result_scan-5.tsv};
\label{MDMB_scan:scan3}

\begin{pgfonlayer}{foreground layer}
	\node[draw,fill=white,draw=black,cap=round, fill opacity=0.95] at (axis cs: -88, 16.5) [anchor=north west] {\begin{tabular}{ccl}
			\ref{MDMB_scan:target} & \quad & \small{target} \\
			\ref{MDMB_scan:bound} & & \small{bound} \\
            \ref{MDMB_scan:MObest} & & \small{envelope}
	\end{tabular}};
\end{pgfonlayer}

\begin{pgfonlayer}{foreground layer}
	\node[draw,fill=white,draw=black,cap=round, fill opacity=0.95] at (axis cs: 88, 16.5) [anchor=north east] {\begin{tabular}{ccl}
            \multicolumn{3}{c}{array gain, $\theta_0$} \\ \hline
			\addlinespace[0.2em] \ref{MDMB_scan:scan1} & \quad & \small{$-10^{\circ}$} \\
			\ref{MDMB_scan:scan2} & & \small{$0^{\circ}$} \\
			\ref{MDMB_scan:scan3} & & \small{$15^{\circ}$}
	\end{tabular}};
\end{pgfonlayer}

\end{axis}
\end{tikzpicture}%
    \caption{Array gain patterns with three different scan directions, scan gain envelope and its bound. }
    \label{fig:MDMB_result_scan}
\end{figure}

The target, fundamental bound, and the best MO result are illustrated in Fig.~\ref{fig:MDMB_result_scan}. The figure shows the array realized gain patterns of the best MO result when feeding the elements with optimal feeding coefficients, and focusing the beam in three different directions. In addition, the scan gain envelope curve is illustrated. The target, which essentially consists of EEPs, is scaled so that it can be compared against the envelope curves. 

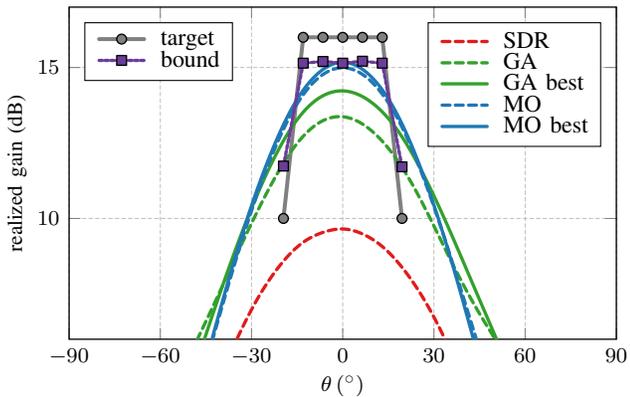
\begin{figure}[t]
    \begin{tikzpicture}

\begin{axis}[%
width=\columnwidth,
height=6cm,
at={(0cm,0cm)},
xmin=-90,
xmax=90,
xlabel={$\theta \, (^{\circ})$},
ymin=6,
ymax=17,
ylabel={realized gain (dB)},
ylabel shift = -0.1cm,
xlabel shift = -0.1cm,
xtick = {-90, -60, ..., 90},
xmajorgrids,
ymajorgrids,
grid style={dashed, cap=round, dash pattern = on 0.5mm off 0.375mm, line width=1/4, opacity=0.5},
axis background/.style={fill=white},
axis line style={solid, line width=0.5pt},
]

\addplot[target] table[]{Figures/multi-driven-multi-beam/MDMB_result_data/MDMB_result-7.tsv};
\label{MDMB:target}


\addplot[SDR, densely dashed] table[]{Figures/multi-driven-multi-beam/MDMB_result_data/MDMB_result-2.tsv};
\label{MDMB:SDR}

\addplot[GA, densely dashed] table[]{Figures/multi-driven-multi-beam/MDMB_result_data/MDMB_result-3.tsv};
\label{MDMB:GA}

\addplot[GA] table[]{Figures/multi-driven-multi-beam/MDMB_result_data/MDMB_result-4.tsv};
\label{MDMB:GAbest}

\addplot[ManuOpt, densely dashed] table[]{Figures/multi-driven-multi-beam/MDMB_result_data/MDMB_result-6.tsv};
\label{MDMB:MO}

\addplot[ManuOpt] table[]{Figures/multi-driven-multi-beam/MDMB_result_data/MDMB_result-5.tsv};
\label{MDMB:MObest}

\addplot[bound, densely dotted] table[]{Figures/multi-driven-multi-beam/MDMB_result_data/MDMB_result-1.tsv};
\label{MDMB:bound}

\begin{pgfonlayer}{foreground layer}
	\node[draw,fill=white,draw=black,cap=round, fill opacity=0.95] at (axis cs: -85, 16.5) [anchor=north west] {\begin{tabular}{ccl}
			\ref{MDMB:target} & \quad & \small{target} \\
			\ref{MDMB:bound} & & \small{bound}
	\end{tabular}};
\end{pgfonlayer}

\begin{pgfonlayer}{foreground layer}
	\node[draw,fill=white,draw=black,cap=round, fill opacity=0.95] at (axis cs: 85, 16.5) [anchor=north east] {\begin{tabular}{ccl}
			\ref{MDMB:SDR} & \quad & \small{SDR} \\
			\ref{MDMB:GA} & & \small{GA} \\
			\ref{MDMB:GAbest} & & \small{GA best} \\
			\ref{MDMB:MO} & & \small{MO} \\
			\ref{MDMB:MObest} & & \small{MO best}
	\end{tabular}};
\end{pgfonlayer}

\end{axis}
\end{tikzpicture}%
    \caption{Scan gain envelopes of the multi-driven antenna optimized using different methods. }
    \label{fig:MDMB_result}
\end{figure}

Figure~\ref{fig:MDMB_result} shows the scan gain envelope curves obtained using different optimization methods. The MO produces the best solutions. The difference to the best GA result is more significant here, emphasizing the superiority of the MO. The realizable SDR result is non-satisfactory, similar to the previous tests.

\begin{figure}[t]
    \centering
    \includegraphics[width=1\linewidth]{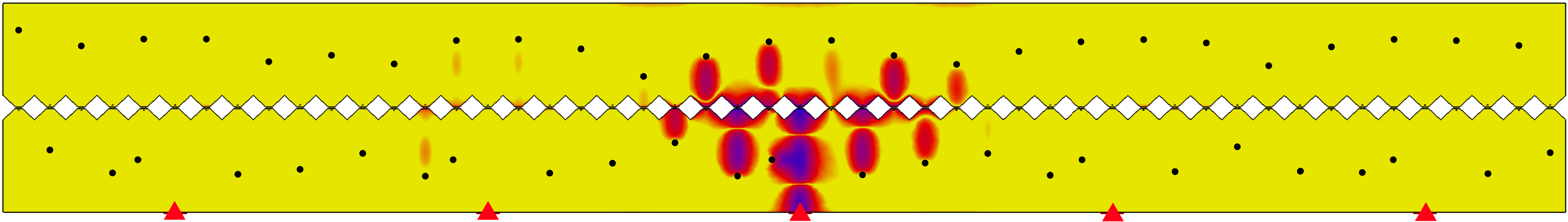}
    \caption{Simulation model of the resulting antenna array. Electric field magnitudes at the substrate are illustrated when the driven port 3 is excited. The terminations are computed using the best MO result.}
    \label{fig:MDMB_antenna_result}
\end{figure}

The best MO solution is implemented in CST. In Fig.~\ref{fig:MDMB_antenna_result}, the electric field distribution between the copper layers is visualized when the port~3 is excited. The black dots illustrate the shorting vias of the scatterer port termination lines. The field distribution expands to a wider area than reserved for a single element. This is beneficial because the larger effective aperture enables the shaping of radiation patterns with more degrees of freedom. 

Figure~\ref{fig:MDMB_antenna_result} also shows that the resulting array is not periodic as the lengths of the transmission lines are different around each driven port. If unit-cell simulations were used, the result would be periodic. Instead, performing the full-array simulation and optimization over each scatterer port termination gives better control over the whole array, taking the edge-element effects into account.

Figure~\ref{fig:MDMB_result_simu} compares the simulation-validated and analytically computed results. The final EEPs of driven ports~1 and~3 are illustrated. The simulation matches the analytical model well except for the back-lobe direction. The radiation in the back direction is stronger in the simulation result because the additional radiation from the microstrip endings, as well as the modified surface waves and losses in the transmission lines, have been taken into account. The EEPs of the two ports are very similar because the individual EEPs are subject to optimization, not the sum of the EEPs.

\begin{figure}[t]
    \begin{tikzpicture}

\begin{polaraxis}[
    width=\linewidth-1.5cm,
    height=\linewidth-1.5cm,
    xticklabel=$\pgfmathprintnumber{\tick}^\circ$,
    xlabel=$\theta$,
    xlabel style = {at={(axis cs:15, 11)}, rotate=45},
    ylabel = realized gain (dB),
    ylabel style = {at={(axis cs:295, 20)}, anchor=center, rotate=90},
    xtick={0,30,...,330},
    ytick={-20, -10, 0, 10},
    ymin=-20, ymax=10,
    y coord trafo/.code=\pgfmathparse{#1+20},
    rotate=-90,
    y coord inv trafo/.code=\pgfmathparse{#1-20},
    x dir=reverse,
    xticklabel style={anchor=-\tick-90},
    yticklabel style={anchor=east, xshift=-3.9cm},
    y axis line style={yshift=-3.9cm},
    ytick style={yshift=-3.9cm},
    grid style={dashed, cap=round, dash pattern = on 0.5mm off 0.375mm, line width=1/4, opacity=0.5},
    legend style={at={(0.72,0)}, anchor=south}
    ]

\addplot[data cs = polarrad, resultEEP2] table[col sep=comma]{Figures/multi-driven-multi-beam/MDMB_result_simu_data/MDMB_polardata_analytical_1.txt};
\label{MDMB_simu:analytical1}

\addplot[data cs = polarrad, resultEEP2, densely dashed] table[col sep=comma]{Figures/multi-driven-multi-beam/MDMB_result_simu_data/MDMB_polardata_simu_1.txt};
\label{MDMB_simu:simu1}

\addplot[data cs = polarrad, resultEEP3] table[col sep=comma]{Figures/multi-driven-multi-beam/MDMB_result_simu_data/MDMB_polardata_analytical_3.txt};
\label{MDMB_simu:analytical3}

\addplot[data cs = polarrad, resultEEP3, densely dashed] table[col sep=comma]{Figures/multi-driven-multi-beam/MDMB_result_simu_data/MDMB_polardata_simu_3.txt};
\label{MDMB_simu:simu3}

\begin{pgfonlayer}{foreground layer}
	\node[draw,fill=white,draw=black,cap=round, fill opacity=0.95] at (axis cs: 265, 25) [anchor=north west] {\begin{tabular}{ccl}
			\ref{MDMB_simu:analytical1} & \quad & \small{analytical 1} \\
			\ref{MDMB_simu:simu1} & & \small{simulated 1} \\
            \ref{MDMB_simu:analytical3} & \quad & \small{analytical 3} \\
			\ref{MDMB_simu:simu3} & & \small{simulated 3}
	\end{tabular}};
\end{pgfonlayer}
        
\end{polaraxis}

\end{tikzpicture}%
    \caption{Analytically computed and simulated EEPs of driven ports 1 and 3. }
    \label{fig:MDMB_result_simu}
\end{figure}
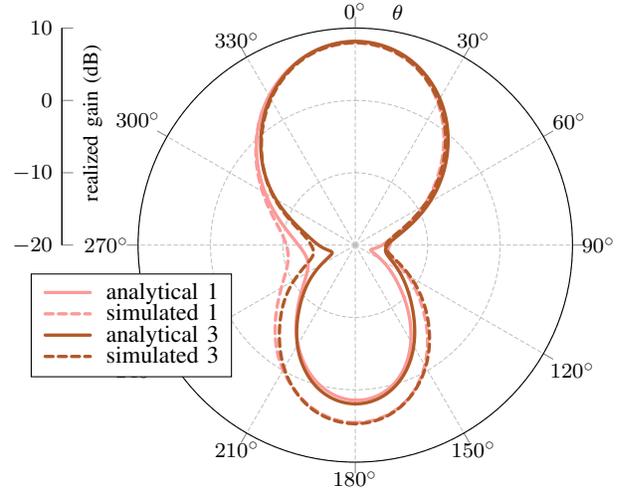

\begin{figure}[t]
    \begin{tikzpicture}

\begin{axis}[%
width=\columnwidth,
height=6cm,
at={(0cm,0cm)},
xmin=4,
xmax=6,
xlabel={frequency (GHz)},
ymin=-15,
ymax=0,
ylabel={$|s_{ii}|$ (dB)},
ylabel shift = -0.2cm,
xlabel shift = -0.1cm,
xmajorgrids,
ymajorgrids,
grid style={dashed, cap=round, dash pattern = on 0.5mm off 0.375mm, line width=1/4, opacity=0.5},
axis background/.style={fill=white},
axis line style={solid, line width=0.5pt},
]

\addplot[SDR] table[]{Figures/multi-driven-multi-beam/MDMB_result_sparameters_data/MDMB_result_sparameters-1.tsv};
\label{MDMB_sparams:s11}

\addplot[ManuOpt] table[]{Figures/multi-driven-multi-beam/MDMB_result_sparameters_data/MDMB_result_sparameters-2.tsv};
\label{MDMB_sparams:s22}

\addplot[GA] table[]{Figures/multi-driven-multi-beam/MDMB_result_sparameters_data/MDMB_result_sparameters-3.tsv};
\label{MDMB_sparams:s33}

\addplot[MPSDR] table[]{Figures/multi-driven-multi-beam/MDMB_result_sparameters_data/MDMB_result_sparameters-4.tsv};
\label{MDMB_sparams:s44}

\addplot[resultEEP2] table[]{Figures/multi-driven-multi-beam/MDMB_result_sparameters_data/MDMB_result_sparameters-5.tsv};
\label{MDMB_sparams:s55}


\begin{pgfonlayer}{foreground layer}
	\node[draw,fill=white,draw=black,cap=round, fill opacity=0.95] at (axis cs: 4.1, -14) [anchor=south west] {\begin{tabular}{ccl}
			\ref{MDMB_sparams:s11} & \quad & \small{$s_{11}$} \\
			\ref{MDMB_sparams:s22} & & \small{$s_{22}$} \\
			\ref{MDMB_sparams:s33} & & \small{$s_{33}$} \\
			\ref{MDMB_sparams:s44} & & \small{$s_{44}$} \\
			\ref{MDMB_sparams:s55} & & \small{$s_{55}$}
	\end{tabular}};
\end{pgfonlayer}

\end{axis}
\end{tikzpicture}%
    \caption{Simulated S-parameters of the optimally loaded antenna array.}
    \label{fig:MDMB_result_sparameters}
\end{figure}
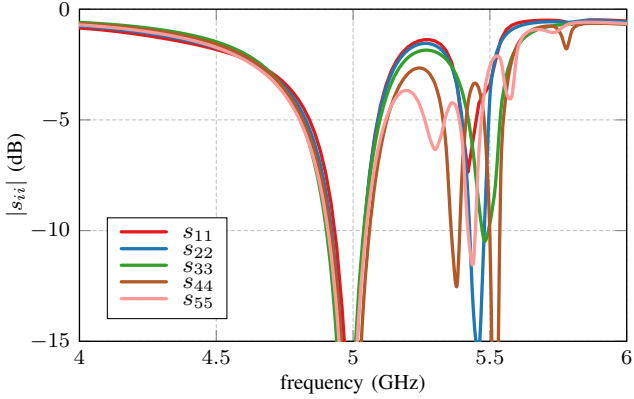

The simulated reflection coefficients of the driven ports, that is, $s_{ii}$ parameters, are illustrated in Fig.~\ref{fig:MDMB_result_sparameters}. Since the realized gain patterns are optimized through the EEPs, rather than directivity patterns, the impedance matching is improved as well. However, due to the point-frequency optimization, the bandwidth is relatively low, approximately 2\,\%.

\section{Discussion}
\label{sec:discussion}

\subsection{Result analysis}

\begin{table}[t]
    \centering
    \caption{Cost function values after optimization using different methods and targets.}
    \begin{threeparttable}
        \begin{tabular}{ccccccc}
            Method & \multicolumn{2}{c}{SDSB}& SDMB & \multicolumn{2}{c}{MDSB} & MDMB \\
            \addlinespace[0.3em] Target $\theta_0$ ($^{\circ}$) & \multicolumn{1}{c}{0} & \multicolumn{1}{c}{20} & \multicolumn{1}{c}{$\left[ -60, 60 \right]$} & \multicolumn{1}{c}{0} & \multicolumn{1}{c}{20} & \multicolumn{1}{c}{$\left[ -19.5, 19.5 \right]$} \\[0.3em] \toprule
            \addlinespace[0.2em] Bound ($\mathrm{V}^2 / \mathrm{m}^2$)& \textbf{77} & \textbf{87} & \textbf{34} & \textbf{247} & \textbf{274} & \textbf{43}\\[0.2em] \hline
            \addlinespace[0.1em] SDR & 1.60 & 1.21 & 1.61 & 2.43 & 2.19 & 5.15\\
            MP-SDR \cite{Corcoles2015} & 1.88 & 1.34 &  & 1.76 & 1.57 &   \\ \hline
            \addlinespace[0.1em] MO, best & \textbf{1.06}$^{\ddagger}$ & \textbf{1.01}$^{\dagger}$ & \textbf{1.01}$^{\dagger}$ & \textbf{1.57}$^{\ddagger}$ & \textbf{1.44}$^{\ddagger}$ & \textbf{1.86} \\
            MO, expected & 1.29 & 1.07 & \textbf{1.01} & 1.60 & 1.49 & 1.98 \\
            MO, worst & 1.41 & 1.50 & 1.04 & 1.65 & 1.52 & 2.97 \\ \hline
            \addlinespace[0.1em] GA, best & \textbf{1.06} & \textbf{1.01} & 1.08 & 1.62 & 1.52 & 2.37 \\
            GA, expected & 1.41 & \textbf{1.01} & 1.20 & 1.67 & 1.55 & 2.46 \\
            GA, worst & 1.41 & 1.34 & 1.24 & 1.75 & 1.67 & 3.06\\ \bottomrule
        \end{tabular}
        \begin{tablenotes}
            \item[$\dagger$] SDR solution used as the initial guess.
            \item[$\ddagger$] MP-SDR solution used as the initial guess.
        \end{tablenotes}
    \end{threeparttable}
    \label{tab:costs}
\end{table}

The final cost function values resulting from different optimization methods are listed in Table~\ref{tab:costs}. The expected results of MO and GA are the solutions where the final cost function value is the median of all 10~solutions. The bound is computed using SDR and is not realizable. The result denoted as the SDR result is realizable, and is extracted from the bound as detailed in Appendix~\hyperref[app:B]{B}.

The bound results are given as cost function values. The numbers describe the maximum difference of squared magnitudes of target EEPs and resulting EEPs, that is, the parameter $t$ in \eqref{eq:SDR_MMF} or \eqref{eq:manopt_MMF}. The other values, which refer to realizable results, are the cost function values relative to the corresponding bound. The relative value of one indicates that the bound result can be realized. The best realizable results are highlighted in bold font.

The best solutions to the single-driven problems are very close to the fundamental bound, with differences of 6\% or even 1\%. Therefore, the local solution is very close to, or potentially exactly at, the global optimum. Both MO and GA provide excellent results in single-driven cases. By comparing these cost function values to Figures \ref{fig:SDSB_result} and \ref{fig:SDMB_result} we see that when the relative cost is below 20\% from the bound, the result and the bound curves are visually very similar. The differences between a few percentage units in relative cost are negligible.

In multi-driven problems, the local solutions are further from the bound, with MO consistently yielding the best results. Furthermore, the expected MO results are superior to the best GA results, especially in the MDMB case.

The MP-SDR formulation results in relatively poor cost-function values but still high realized gain. For example, the MP-SDR-optimized broadside-radiating multi-driven antenna has a cost value of 1.76, but, as shown in Fig.~\ref{fig:MDSB_result}, the main-beam gain is practically equivalent to that of the best MO result with the cost of 1.57. This indicates that the EEPs resulting from MP-SDR are non-identical and the high gain is produced by only a couple of elements. The cost values indicate the maximum difference between the target and the result, considering all driven ports and all desired directions. 

In the test cases, we use the realizable SDR and the MP-SDR solutions, along with random ones, as initial guesses for the local search algorithm. In MDSB cases, the best realizable result is obtained by the MO with the MP-SDR solution as the initial guess. Additionally, these tests indicate that the SDR solution is also a good starting point for MO. However, starting from either the SDR solution or the MP-SDR solution does not guarantee convergence to the best result. Therefore, random starting points are also needed.

Although, in this work, the SDR is primarily used to compute the fundamental bound, we also try to extract a realizable solution based on the bound. The straightforward extraction method is presented in Appendix~\hyperref[app:B]{B}. The realizable SDR results are underperforming. A more sophisticated extraction using a heuristic approach, as done in \cite{Fuchs2014} and \cite{Liu2020}, could provide better realizable SDR results which could further improve the initial guess for the local search algorithms. Note that, the extraction problem itself is non-convex because if it were not, the SDR approach would solve the NP-hard problem in polynomial time.

On the other hand, when analyzing the MDSB results, we see that the MP-SDR method offers excellent performance, despite being based on convex relaxation and a very straightforward result extraction. The MP-SDR formulation is simpler, with only $(\ND + \NP)^2$ complex optimization variables and $\NP + 1$ constraints. In contrast, our MMF SDR formulation has $(\ND \NP)^2 + \ND \NP + 1$ variables and $2 \ND \NP + \ND + 1$ constraints in single-beam problems. The increased number of variables and constraints in our method is due to its applicability to the most general multi-driven multi-beam problem, which cannot be solved using the MP-SDR formulation.

\subsection{Computational complexity}
We analyze the computational complexity of the presented optimization framework using the multi-driven antenna shown in Fig.~\ref{fig:array_full}. In this analysis, the optimization goal is the same as in the MDMB study.

The number of considered driven ports $\ND$ goes from one to five. The number of scatterer ports is $\NP \in \left[1, 50 \right]$. The scatterer port terminations are optimized for all combinations of selected driven and scatterer ports, that is, for 250 separate cases. The ports left out of consideration, either driven or scatterer ones, are short-circuited. The considered scatterer ports are selected based on the proximity of the considered driven ports. The selection order of driven and scatterer ports can be found in the MATLAB scripts published as supplementary material.

Figure~\ref{fig:complexity} shows the CPU times required for the convergence of the optimization algorithms. The bound is computed using SDR, and the realizable SDR result is used as the initial guess for the MO algorithm. The MO and GA are run only once per problem.

\begin{figure}[t]
    \begin{tikzpicture}


\begin{axis}[
name={COMP_TIME_N3},
width=4.1cm,
height=4.5cm,
title={$\ND = 3$},
every axis title/.style={below, at={(0.5,1)},draw=black,fill=white},
axis y line*=left,
at={(0,0)},
xmin=5,
xmax=60,
xlabel={$\NP$},
ymin=0,
ymax=300,
ylabel={time (s)},
ylabel shift = -0.15cm,
xlabel shift = -0.1cm,
xtick = {10, 30, 50},
xmajorgrids,
ymajorgrids,
grid style={dashed, cap=round, dash pattern = on 0.5mm off 0.375mm, line width=1/4, opacity=0.5},
axis background/.style={fill=white},
axis line style={solid, line width=0.5pt},
]

\addplot[line width=1.25pt, PairedJ, cap=round, join=round] table[col sep=comma]{Figures/complexity/t_sdr_N3.txt};
\label{COMP_TIME_N3:SDR}

\addplot[ManuOpt] table[col sep=comma]{Figures/complexity/t_mo_N3.txt};
\label{COMP_TIME_N3:MO}

\addplot[GA] table[col sep=comma]{Figures/complexity/t_ga_N3.txt};
\label{COMP_TIME_N3:GA}

\addplot[line width=1.75pt, gray, cap=round, join=round, smooth] table[col sep=comma]{Figures/complexity/t_tot_N3.txt};
\label{COMP_TIME_N3:TOT}

\addplot[resultEEP2, line width=0.75pt] table[col sep=comma]{Figures/complexity/t_fit_N3.txt};
\label{COMP_TIME_N3:FIT}

\end{axis}






\begin{axis}[%
name={COMP_TIME_N4},
width=4.1cm,
height=4.5cm,
title={$\ND = 4$},
every axis title/.style={below, at={(0.5,1)},draw=black,fill=white},
yticklabel=\empty,
axis y line*=left,
at={(COMP_TIME_N3.right of south east), anchor=right of south west},
xshift=0.1cm,
xmin=5,
xmax=60,
xlabel={$\NP$},
ymin=0,
ymax=300,
xlabel shift = -0.1cm,
xtick = {10, 30, 50},
xmajorgrids,
ymajorgrids,
grid style={dashed, cap=round, dash pattern = on 0.5mm off 0.375mm, line width=1/4, opacity=0.5},
axis background/.style={fill=white},
axis line style={solid, line width=0.5pt},
]

\addplot[line width=1.25pt, PairedJ, cap=round, join=round] table[col sep=comma]{Figures/complexity/t_sdr_N4.txt};
\label{COMP_TIME_N4:SDR}

\addplot[ManuOpt] table[col sep=comma]{Figures/complexity/t_mo_N4.txt};
\label{COMP_TIME_N4:MO}

\addplot[GA] table[col sep=comma]{Figures/complexity/t_ga_N4.txt};
\label{COMP_TIME_N4:GA}

\addplot[line width=1.75pt, gray, cap=round, join=round, smooth] table[col sep=comma]{Figures/complexity/t_tot_N4.txt};
\label{COMP_TIME_N4:TOT}

\addplot[resultEEP2, line width=0.75pt] table[col sep=comma]{Figures/complexity/t_fit_N4.txt};
\label{COMP_TIME_N4:FIT}




\begin{pgfonlayer}{foreground layer}
	\node[draw,fill=white,draw=black,cap=round, fill opacity=0.95] at (axis cs: 27.5, 315) [anchor=south] {\begin{tabular}{cccccccccc}
			\addlinespace[-0.2em]\ref{COMP_TIME_N4:SDR} & bound \quad & \ref{COMP_TIME_N4:MO} & MO \quad & \ref{COMP_TIME_N4:TOT} & bound + MO \quad & \ref{COMP_TIME_N4:FIT} & fitted \quad & \ref{COMP_TIME_N4:GA} & GA \\[-0.6em]
	\end{tabular}};
\end{pgfonlayer}

\end{axis}






\begin{axis}[%
name={COMP_TIME_N5},
width=4.1cm,
height=4.5cm,
title={$\ND = 5$},
every axis title/.style={below, at={(0.5,1)},draw=black,fill=white},
yticklabel=\empty,
axis y line*=left,
at={(COMP_TIME_N4.right of south east), anchor=right of south west},
xshift=0.1cm,
xmin=5,
xmax=60,
xlabel={$\NP$},
ymin=0,
ymax=300,
xlabel shift = -0.1cm,
xtick = {10, 30, 50},
xmajorgrids,
ymajorgrids,
grid style={dashed, cap=round, dash pattern = on 0.5mm off 0.375mm, line width=1/4, opacity=0.5},
axis background/.style={fill=white},
axis line style={solid, line width=0.5pt},
]

\addplot[line width=1.25pt, PairedJ, cap=round, join=round] table[col sep=comma]{Figures/complexity/t_sdr_N5.txt};
\label{COMP_TIME_N5:SDR}

\addplot[ManuOpt] table[col sep=comma]{Figures/complexity/t_mo_N5.txt};
\label{COMP_TIME_N5:MO}

\addplot[GA] table[col sep=comma]{Figures/complexity/t_ga_N5.txt};
\label{COMP_TIME_N5:GA}

\addplot[line width=1.75pt, gray, cap=round, join=round, smooth] table[col sep=comma]{Figures/complexity/t_tot_N5.txt};
\label{COMP_TIME_N5:TOT}

\addplot[resultEEP2, line width=0.75pt] table[col sep=comma]{Figures/complexity/t_fit_N5.txt};
\label{COMP_TIME_N5:FIT}

\end{axis}






\end{tikzpicture}%
    \caption{Computing times of problems with $\ND$ driven ports and $\NP$ scatterer ports. The fitted polynomial approximates the computational complexity of obtaining the bound and a realizable solution.}
    \label{fig:complexity}
\end{figure}

The most time-consuming part of the optimization framework is computing the fundamental bound. Nevertheless, the bound must be computed only once for a given initial design, whereas the local search algorithms often require multiple runs to find a satisfactory solution. The MO and GA algorithms converge almost linearly. Their computing times are comparable, although GA is slightly faster. However, according to the tests reported in Table~\ref{tab:costs}, the MO algorithm generally provides better results than GA.

In our problem setting, the total computation time (bound and MO) accurately follows a polynomial of order
\begin{align}
    p(\ND, \NP) = \mathcal{O}(\max(\NP^4, \ND^2 \NP^2) ).
\end{align}
The polynomial is fitted to the measured time in Fig.~\ref{fig:complexity}. This empirically found complexity of the fourth-degree polynomial of a semi-definite program is in line with \cite{Luo2010}. 

Extrapolation of the polynomial gives estimates of the time requirements of large-scale problems. For instance, with \mbox{$\ND = 20$} and \mbox{$\NP = 200$}, the computing time would be $2.9$~days. Note that, this estimate is valid only for the studied antenna topology, and further analysis is required to generalize the complexity estimate. The computation could be significantly accelerated using parallel computing or by reducing the problem size with unit-cell modeling.

\section{Conclusion}
\label{sec:conclusion}
This paper introduced a method for computing a tight fundamental bound for the problem of optimizing reactively loaded antenna arrays. The bound served as a useful benchmark to evaluate the performance of local solutions concerning the non-convex optimization problem. However, because the bound was the solution to the relaxed problem, it could not be realized in practice. Therefore, other optimization methods were necessary for realization.

The work further employed the Riemannian augmented Lagrangian method as a local search algorithm for tackling this problem. Comparisons with state-of-the-art algorithms, including the genetic algorithm and the minimum-power semi-definite relaxation method, demonstrated the superiority of the RALM. The algorithms developed in this work were made freely accessible through MATLAB codes provided as supplementary material.

The MP-SDR formulation proved particularly effective in single-beam problems, where matching circuits could be added to the antenna elements, and amplitude-tapered feeding coefficients could be applied to accommodate non-identical embedded element patterns. In contrast, the manifold optimization produced identical EEPs, which were advantageous for practical applications. Also, MO allowed for the consideration of impedance matching. While the genetic algorithm was computationally fast, it yielded inferior solutions compared to MO.

The current SDR and MO algorithms were developed for point frequencies only, and therefore the results were highly frequency-selective. In the state-of-the-art, GA would be the most suitable algorithm for wideband design, as demonstrated in \cite{Lamey2021}. Developing the MO algorithm for wideband antennas is considered future work. In addition, a deeper analysis of the tightness of the semi-definite relaxation would be useful in the future.

The demonstrations with the connected bowtie-slot antenna validated the usability of the proposed framework for applications involving antennas with single or multiple excitations, targeting both beam-focusing and beam-shaping objectives. Furthermore, the simulation results confirmed the practical applicability of the framework.


\section*{Acknowledgment}
The research utilized the Aalto Electronics-ICT infrastructure of Aalto University.


\ifCLASSOPTIONcaptionsoff
  \newpage
\fi

\section*{Appendix A: Manifold optimization}
\label{app:A}

\subsection{Problem formulation}
Consider the optimization problem~\eqref{eq:manopt_MMF}. Recall that the indices $n$ and $l$ refer to a driven port and a beam direction~$(\theta_l, \varphi_l)$, respectively. The driven port's EEP is denoted by $\newe_{nl}(\mr)$. The $\NP$ scatterer ports are terminated by reactances with reflection coefficients $\mr \in \mathbb{C}^{\NP}$. The optimization minimizes the maximum distance between the pre-defined target EEP magnitudes $|\tilde{e}_{nl}|^2$ and $|\newe_{nl}(\mr)|^2$ considering all driven ports $n$ and beam directions $l$.

The problem \eqref{eq:manopt_MMF} can be rewritten as
\begin{mini}|l|[1]
    {\bm{x} \in \mathcal{M}}
    {x_1}
    {\label{eq:manopt_MMF_mod}}
    {}
    \addConstraint{g_{nlk}(\bm{x})}{\leq 0, \quad}{\forall n \in [1, \ND], l \in [1, L], k \in \{1,2\}}.
\end{mini}
The vector $\bm{x} = [ t \;\; \mr^{\mathrm{T}}]^{\mathrm{T}}$ combines the parameters $t$ and $\bm{r}$, and the first entry of $\bm{x}$ is $x_1 = t$. Therefore, we need to introduce the product manifold
\begin{align}
    \mathcal{M} = \mathbb{R} \times \mathcal{C}^{\NP} = \big\{ &\bm{x} \in \mathbb{C}^{\NP+1} : \Im(x_1) = 0, \nonumber \\
    & |x_i| = 1, \forall i = 2, \ldots, \NP+1 \big\}.
\end{align}

The inequality constraint functions are formed from the constraints~$-t \leq |\newe_{nl}(\mr)|^2 - |\tilde{e}_{nl}|^2 \leq t$ as follows:
\begin{align}
    g_{nlk}(\bm{x}) = \begin{cases}
        |\newe_{nl}(\bm{x})|^2 - |\tilde{e}_{nl}|^2 - x_1, & k = 1, \\
        -x_1 - |\newe_{nl}(\bm{x})|^2 + |\tilde{e}_{nl}|^2, & k = 2.
    \end{cases}
\end{align}
The constraints limit the difference between the resulting EEP magnitudes and target magnitudes within the limit $x_1 = t$. 

The resulting EEPs can be computed based on~\eqref{eq:maPn} and~\eqref{eq:new_enl} as
\begin{align}
    \newe_{nl}(\bm{x}) = & \eD_{nl} + (\bm{s}^{\mathrm{PD}}_n)^{\mathrm{T}} \bm{M}(\bm{x})^{-1} \meP_l,
\end{align}
where
\begin{align}
    \bm{M}(\bm{x}) = \diag([x_2, \ldots, x_{\NP+1}])^{-1} - \SPP,
\end{align}
and $\bm{s}^{\mathrm{PD}}_n$ is the $n$-th column of the $\SPD$. 

The Riemannian augmented Lagrangian method minimizes the Lagrangian function with respect to the primal variable~$\bm{x}$ and dual variable~$\bm{\lambda}$, as described in Algorithm 1 of~\cite{Liu2020}. The augmented Lagrangian function of the problem~\eqref{eq:manopt_MMF_mod} is
\begin{align}
    \mathcal{L}_{\rho}(\bm{x}, \bm{\lambda}) = x_1 + \frac{\rho}{2} \sum_{n,l,k} \max \bigg\{ 0, \frac{\lambda_{nlk}}{\rho} + g_{nlk}(\bm{x})\bigg\}^2,
\end{align}
where 
\begin{align}
    \sum_{n,l,k} = \sum_{n=1}^{\ND} \sum_{l=1}^{L} \sum_{k=1}^{2},
\end{align}
and $\lambda_{nlk} \geq 0$ are the Lagrangian multipliers, that is, the dual variables. The penalty parameter is $\rho > 0$.

\subsection{Minimization of Lagrangian}
The first step of the RALM minimizes the Lagrangian function~$\mathcal{L}_{\rho}(\bm{x}, \bm{\lambda})$ on the Manifold~$\mathcal{M}$ with respect to~$\bm{x}$. The dual variable~$\bm{\lambda}$ and the penalty parameter~$\rho$ are fixed. After that,~$\bm{\lambda}$ and~$\rho$ are updated as described in Algorithm 1 of~\cite{Liu2020}. These steps are repeated until convergence.

The RBFGS algorithm is used in the first step to minimize~$\mathcal{L}_{\rho}(\bm{x}, \bm{\lambda})$. The algorithm determines the next iteration point based on the steepest descent direction which is computed using the intrinsic Riemannian gradient of~$\mathcal{L}_{\rho}(\bm{x}, \bm{\lambda})$. To obtain the Riemannian gradient, we first derive the Euclidean gradient of the Lagrangian.

The Euclidean partial derivatives of $\mathcal{L}_{\rho}(\bm{x}, \bm{\lambda})$ with respect to $x_i$ are
\begin{align}
    & \frac{\partial \mathcal{L}_{\rho}(\bm{x}, \bm{\lambda})}{ \partial x_i} = \delta_{i1} \nonumber \\
    &+ \rho  \sum_{n,l,k} \max \bigg\{ 0, \frac{\lambda_{nlk}}{\rho} + g_{nlk}(\bm{x})\bigg\} \frac{\partial g_{nlk}(\bm{x})}{\partial x_i},
\end{align}
where $\delta_{ij}$ is the Kronecker delta function and
\begin{align}
    \dfrac{\partial g_{nlk}(\bm{x})}{\partial x_i} = \begin{cases}
        -1, & i = 1 \\
        \begin{cases}
            \dfrac{\partial |\newe_{nl}(\bm{x})|^2}{\partial x_i}, & k = 1 \\
            -\dfrac{\partial |\newe_{nl}(\bm{x})|^2}{\partial x_i}, & k = 2 
        \end{cases}, & i > 1
    \end{cases}.
\end{align}

The partial derivatives of EEP magnitudes with respect to the reflection coefficients, that is, $x_i$ for $i>1$, are
\begin{align}
    \dfrac{\partial|\newe_{nl}(\bm{x})|^2}{\partial x_m} = 2 \big( \alpha_m (\bm{s}_n^{\mathrm{PD}})^{\mathrm{T}} \bm{M}^{-1} \bm{u}_m \bm{u}_m^{\mathrm{T}} \bm{M}^{-1} \meP_l \big)^* \newe_{nl},
\end{align}
where
\begin{align}
    \alpha_m = -\frac{1}{x_m^{\mathrm{Im}} - \ji x_m^{\mathrm{Re}}},
\end{align}
$\ji$ denotes the imaginary unit, and~$(\cdot)^*$ complex conjugation. The real variables~$x_m^{\mathrm{Re}}$ and~$x_m^{\mathrm{Im}}$ are the real and imaginary parts of~$x_m$, respectively. The vector~$\bm{u}_m \in \mathbb{R}^{\NP}$ has 1 in the entry of row~$m$, and the other entries are zero. The index~$m = i-1$ is used to refer to the reflection coefficient part of vector~$\bm{x}$.

Finally, the Euclidean gradient of the Lagrangian with respect to $\bm{x}$ is the vector of partial derivatives:
\begin{align}
    \nabla_{\bm{x}} \mathcal{L}_{\rho}(\bm{x}, \bm{\lambda}) = \begin{bmatrix}
        \dfrac{\partial \mathcal{L}_{\rho}(\bm{x}, \bm{\lambda})}{ \partial x_1}, \ldots, \dfrac{\partial \mathcal{L}_{\rho}(\bm{x}, \bm{\lambda})}{ \partial x_{\NP+1}}
    \end{bmatrix}^{\mathrm{T}}.
\end{align}

The intrinsic Riemannian gradient, used in the optimization algorithm, is computed based on the Euclidean gradient by projecting the Euclidean gradient to the tangent space of the manifold~$\mathcal{M}$. The explicit definitions for the projection operator can be found in~\cite{Alhujaili2019}, for instance. The first term of the Euclidean gradient must not be projected since it is already on the correct manifold.

When the steepest descent direction is found, a line search is performed to examine the length of the step moved towards the descent direction. After moving in the descent direction on the tangent space of the manifold, the resulting point is retracted back onto the manifold~$\mathcal{M}$. The projection of the gradient, line search, and retraction operations are implemented in the Manopt toolbox~\cite{manopt}.

\subsection{Solver parameters}

\begin{table}[t]
    \centering
    \caption{Solver parameter values for RALM.}
    \begin{tabular}{ccc}
        Parameter & Value & Explanation \\ \toprule
        $d_{\min}$ & $10^{-8}$ & Minimum step size \\
        $\rho_0$ & 1 & Starting penalty coefficient\\
        $\theta_{\rho}$ & 3.3 & Penalty coefficient's increment factor \\
        $\theta_{\varepsilon}$ & 0.8 & Accuracy tolerance's increment factor\\
        $\theta_{\sigma}$ & 0.8 & $\sigma$-indicator's increment factor \\
        $\varepsilon_0$ & $10^{-3}$ & Starting accuracy tolerance\\
        $\varepsilon_{\min}$ & $10^{-6}$ & Minimum accuracy tolerance\\
        $\lambda_0$ & 0.1 & Starting multiplier values \\
        $\lambda_{\min}$ & $10^{-4}$ & Minimum multiplier value \\
        $\lambda_{\max}$ & $10^6$ & Maximum multiplier value \\ \bottomrule
    \end{tabular}
    \label{tab:RALM_parameters}
\end{table}

The maximum number of iterations for the RALM solver is set to~1000. The initial guess for Lagrangian multipliers is chosen so that each multiplier has the same value~$\lambda_0$, given in Table~\ref{tab:RALM_parameters}.

When minimizing~$\mathcal{L}_{\rho}(\bm{x}, \bm{\lambda})$ with respect to~$\bm{x}$ using the RBFGS, the maximum number of iterations is set to~200. The minimum number of RBFGS iterations is set to~10. The memory usage is not limited in the RBFGS algorithm. The rest of the RBFGS solver settings are the defaults of the Manopt toolbox \cite{manopt}.

\section*{Appendix B: Semi-definite relaxation}
\setcounter{subsection}{0}
\label{app:B}

Consider the optimization problem \eqref{eq:SDR_MMF}. In this Appendix, we derive the quadratic equality and inequality constraint functions $f_i$, $g_j$, and $h_j$. Because not all functions are positive semi-definite, we apply semi-definite relaxation to relax the problem into a convex form. Then, the convex problem is solved using semi-definite programming.

\subsection{Vectorization}
To construct the quadratic constraint functions, the problem is vectorized. Denote a column $m$ of the matrix $\bm{X}$ as $\bm{x}_m$. The vectorization operator $\vect$ stacks columns of a matrix below each other. The vectorization of a $N \times M$ matrix $\bm{X}$ is
\begin{align}
    \bar{\bm{x}} = \vect(\bm{X}) = \begin{bmatrix}
        \bm{x}_1 \\
        \vdots \\
        \bm{x}_M
    \end{bmatrix} \in \mathbb{C}^{(NM) \times 1}.
\end{align}
We denote vectorized matrices with the bar over a bold small-case letter. In addition, we denote diagonally repeating matrices with capital bold letters with the bar as
\begin{align}
    \bar{\bm{Y}} = \mathbf{I}_N \otimes \bm{Y} = \begin{bmatrix}
        \bm{Y} & \cdots & \bm{0} \\
        \vdots & \ddots & \vdots \\
        \bm{0} & \cdots & \bm{Y}
    \end{bmatrix},
\end{align}
where $\mathbf{I}_N$ is $N$-size identity matrix and $\otimes$ denotes the Kronecker product. 

Let $\newmE$ be a $\ND \times L$ matrix of driven ports' EEPs to all desired directions after terminating the scatterer ports. Based on \eqref{eq:new_enl}, it can be expressed as
\begin{align} \label{eq:newmE}
    \newmE^{\mathrm{T}} = (\mEP)^{\mathrm{T}} \bm{A}^{\mathrm{P}} + (\mED)^{\mathrm{T}},
\end{align}
where $\bm{A}^{\mathrm{P}} \in \mathbb{C}^{\NP \times \ND}$ is the matrix of incident waves into the scatterer ports. A column $n$ of $\bm{A}^{\mathrm{P}}$ contains the incident waves to all scatterer ports when the driven port $n$ is excited with unit magnitude and the other driven ports are terminated to the reference impedances. The incident waves depend on the reflection coefficients of the scatterer ports as expressed in \eqref{eq:maPn}. The matrix $\bm{A}^{\mathrm{P}}$ is the optimization variable.

We vectorize \eqref{eq:newmE} as
\begin{align}
    \hat{\bar{\bm{e}}} = \vect(\newmE^{\mathrm{T}}) &= \vect((\mEP)^{\mathrm{T}} \bm{A}^{\mathrm{P}} + (\mED)^{\mathrm{T}}) \nonumber \\
    &= (\bm{I}_{\ND} \otimes (\mEP)^{\mathrm{T}} ) \vect(\bm{A}^{\mathrm{P}}) + \vect((\mED)^{\mathrm{T}}) \nonumber \\
    &= \bar{\bm{E}}^{\mathrm{P}} \bar{\bm{a}}^{\mathrm{P}} + \bar{\bm{e}}^{\mathrm{D}},
\end{align}
where 
\begin{align}
    \bar{\bm{a}}^{\mathrm{P}} &= \vect( (\mR^{-1} - \SPP)^{-1} \SPD ) \nonumber \\
    &= (\bm{I}_{\ND} \otimes (\mR^{-1} - \SPP)^{-1}) \vect(\SPD) \nonumber \\
    &= (\bm{I}_{\ND} \otimes \mR^{-1} - \bm{I}_{\ND} \otimes \SPP ) ^{-1} \vect(\SPD) \nonumber \\
    &= (\bar{\mR}^{-1} - \bar{\bm{S}}^{\mathrm{PP}} ) ^{-1} \bar{\bm{s}}^{\mathrm{PD}}.
    \label{eq:ap_vec}
\end{align}
The problem is now to find $\bar{\bm{a}}^{\mathrm{P}} \in \mathbb{C}^{\ND \NP}$ that minimizes the maximum magnitude difference between elements of $\hat{\bar{\bm{e}}} \in \mathbb{C}^{\ND L}$ and $\tilde{\bar{\bm{e}}} \in \mathbb{C}^{\ND L}$ where $\tilde{\bar{\bm{e}}} = \operatorname{vec}(\tilde{\bm{E}})$ is the vector of target EEPs. In addition, $\bar{\bm{a}}^{\mathrm{P}}$ must satisfy \eqref{eq:ap_vec} with a constant-modulus constrained and diagonally repeating $\bar{\mR}$. These constraints are considered in the functions $g$ and $h$ derived in the next section.

Let us ease the notation by defining
\begin{align*}
    \bm{x} &= \bar{\bm{a}}^{\mathrm{P}}, &
    \bm{y} &= \hat{\bar{\bm{e}}}, &
    \tilde{\bm{y}} &= \tilde{\bar{\bm{e}}}, \\
    \bm{Q} &= \bar{\bm{E}}^{\mathrm{P}}, &
    \bm{c} &= \bar{\bm{e}}^{\mathrm{D}}, &
    \bm{p} &= \bar{\bm{s}}^{\mathrm{PD}}, \\
    \bm{S} &= \bar{\bm{S}}^{\mathrm{PP}}.
\end{align*}
In addition, let us define a basis vector~$\bm{u}_i$, that has~1 in entry~$i$, and other entries are zero.

\subsection{Optimization constraints}
Let us first derive the functions $f_i, \quad i=1\ldots \ND L$ \eqref{eq:SDR_minimaxconst} which constrain the error between the resulting and target EEP magnitudes within the parameter $t$. Consider the following inequality:
\begin{align} \label{eq:fstart}
    -t \leq |\newe_{nl}|^2 - |\tilde{e}_{nl}|^2 \leq t, \quad \forall n \in [1, N], l \in [1, L],
\end{align}
where $|\tilde{e}_{nl}|^2$ is the given target magnitude. Using the vectorized formulation and the eased notation, \eqref{eq:fstart} is expressed as
\begin{align}
    -t \leq |\bm{u}_i^{\mathrm{T}} \bm{y}|^2 - |\tilde{{y}}_i|^2 \leq t, \quad \forall i \in [1, NL].
\end{align}
The quadratic constraint function $f_i(\bm{x})$ is then
\begin{align}
    f_i(\bm{x}) & = |\bm{u}_i^{\mathrm{T}} \bm{y}|^2 - |\tilde{{y}}_i|^2 \nonumber \\
    & = |\bm{u}_i^{\mathrm{T}} \bm{Q} \bm{x} + \bm{u}_i^{\mathrm{T}} \bm{c}|^2  - |\tilde{{y}}_i|^2 \nonumber \\
    & = \bm{x}^{\mathrm{H}} \bm{Q}^{\mathrm{H}} \bm{u}_i \bm{u}_i^{\mathrm{T}} \bm{Q} \bm{x} + 2 \Re ( \bm{c}^{\mathrm{H}} \bm{u}_i \bm{u}_i^{\mathrm{T}} \bm{Q} \bm{x})  + |c_i|^2  - |\tilde{{y}}_i|^2.
\end{align}

Next, we derive formulas for the equality constraint functions $g_j, \quad j=1\ldots \ND \NP$ in \eqref{eq:SDR_passivityconst}. These ensure that the incident waves $\bm{A}^{\mathrm{P}}$ can be realized when the reflection coefficients of the scatterer ports have unit magnitudes, that is, $|r_m|~=~1,~\quad~\forall~m\in~[1,~\NP]$. Based on \eqref{eq:ap_vec}, the constant modulus constraints can be transformed to quadratic constraints in $\bm{x}$, as
\begin{align}
    \bm{x} = (\bar{\mR}^{-1} - \bm{S} )^{-1} \bm{p} \quad \Leftrightarrow \quad \bar{\mR} (\bm{S} \bm{x} + \bm{p}) = \bm{x}.
    \label{eq:x_vec_ap}
\end{align}

If the vector $\bm{x}$ satisfies the condition that the element-wise magnitudes of $\bm{S} \bm{x} + \bm{p}$ and $\bm{x}$ are equal, then we can form a matrix $\bar{\mR}$ that is diagonal and has unit-magnitude entries. Thus, the reactivity constraints $g_j$ are
\begin{align}
    g_j(\bm{x}) & = |\bm{u}_j^{\mathrm{T}} \bm{S} \bm{x} + \bm{u}_j^{\mathrm{T}} \bm{p} |^2 - |\bm{u}_j^{\mathrm{T}} \bm{x}|^2 \nonumber \\
    & = \bm{x}^{\mathrm{H}} (\bm{S}^{\mathrm{H}} \bm{u}_j \bm{u}_j^{\mathrm{T}} \bm{S} - \bm{u}_j \bm{u}_j^{\mathrm{T}})\bm{x} \nonumber \\
    & + 2 \Re (\bm{p}^{\mathrm{H}} \bm{u}_j \bm{u}_j^{\mathrm{T}} \bm{S} \bm{x} ) + |p_j|^2.
\end{align}

Next, we derive the quadratic equality constraint functions~$h_j, \quad j=1\ldots \ND \NP$ in~\eqref{eq:SDR_independenceconst}. The reactivity constraints $g_j$ are insufficient to require that $\mR$ is repeating in the diagonal of $\bar{\mR}$. Additional constraints are required for satisfying $\bar{\mR} = \bm{I}_N \otimes \mR$. We require that the diagonal elements of $\bar{\mR}$ satisfy $\bar{r}_j = \bar{r}_k, \quad \forall j \in [1, NM]$, where
\begin{align}
    k = j - \NP (\lceil j / \NP \rceil - 1),
\end{align}
and $\lceil \cdot \rceil$ is the ceiling operator. Based on \eqref{eq:x_vec_ap}, the diagonal elements of $\bar{\mR}$ can be expressed as
\begin{align}
    \bar{r}_j = \frac{\bm{u}_j^{\mathrm{T}} \bm{x}}{\bm{u}_j^{\mathrm{T}} \bm{S} \bm{x} + \bm{u}_j^{\mathrm{T}} \bm{p}}.
    \label{eq:bar_rj}
\end{align}

The reactivity constraints $g_j$ ensure constant modulus for~$\bar{r}_j$, and consequently it holds that~$\bar{r}_j = 1 / \bar{r}_j^*$. To formulate the problem more suitable for the solver algorithm, we derive the repetition constraints $h_j$ by requiring that
\begin{align}
    \bar{r}_j &= \frac{1}{\bar{r}_k^*} \nonumber \\
    \Rightarrow \frac{\bm{u}_j^{\mathrm{T}} \bm{x}}{\bm{u}_j^{\mathrm{T}} \bm{S} \bm{x} + \bm{u}_j^{\mathrm{T}} \bm{p}} &= \frac{\bm{u}_k^{\mathrm{T}} \bm{S}^* \bm{x}^* + \bm{u}_k^{\mathrm{T}} \bm{p}^*}{\bm{u}_k^{\mathrm{T}} \bm{x}^*}.
\end{align}
Consequently, the repetition constraint functions $h_j$ are
\begin{align}
    h_j(\bm{x}) = &  \bm{x}^{\mathrm{H}} (\bm{S}^{\mathrm{H}} \bm{u}_k \bm{u}_j^{\mathrm{T}} \bm{S} - \bm{u}_k \bm{u}_j^{\mathrm{T}}) \bm{x} \nonumber \\
    & + \bm{p}^{\mathrm{H}} \bm{u}_k \bm{u}_j^{\mathrm{T}} \bm{S} \bm{x} + \bm{x}^{\mathrm{H}} \bm{S}^{\mathrm{H}} \bm{u}_k \bm{u}_j^{\mathrm{T}} \bm{p} + p_k^* p_j.
\end{align}

\subsection{Relaxation}
Quadratic terms can be expressed as $\bm{x}^{\mathrm{H}} \bm{F} \bm{x} = \trace(\bm{F} \bm{x} \bm{x}^{\mathrm{H}})$. The terms become linear when the variable vector $\bm{x}$ is replaced by the matrix $\bm{X} = \bm{x} \bm{x}^H$. We transform the quadratic constraint functions of the problem into linear forms as follows:
\begin{align}
    f_i(\bm{x}, \bm{X}) = & \trace( \bm{Q}^{\mathrm{H}} \bm{u}_i \bm{u}_i^{\mathrm{T}} \bm{Q} \bm{X}) \nonumber \\
                          & + 2 \Re ( \bm{c}^{\mathrm{H}} \bm{u}_i \bm{u}_i^{\mathrm{T}} \bm{Q} \bm{x})  + |c_i|^2  - |\tilde{{y}}_i|^2 \\
    g_j(\bm{x}, \bm{X}) = & \trace( (\bm{S}^{\mathrm{H}} \bm{u}_j \bm{u}_j^{\mathrm{T}} \bm{S} - \bm{u}_j \bm{u}_j^{\mathrm{T}}) \bm{X}) \nonumber \\
                          & + 2 \Re (\bm{p}^{\mathrm{H}} \bm{u}_j \bm{u}_j^{\mathrm{T}} \bm{S} \bm{x} ) + |p_j|^2 \\
    h_j(\bm{x}, \bm{X}) = & \trace( (\bm{S}^{\mathrm{H}} \bm{u}_k \bm{u}_j^{\mathrm{T}} \bm{S} - \bm{u}_k \bm{u}_j^{\mathrm{T}}) \bm{X}) \nonumber \\
                          & + \bm{p}^{\mathrm{H}} \bm{u}_k \bm{u}_j^{\mathrm{T}} \bm{S} \bm{x} + \bm{x}^{\mathrm{H}} \bm{S}^{\mathrm{H}} \bm{u}_k \bm{u}_j^{\mathrm{T}} \bm{p} + p_k^* p_j.
\end{align}
With the constraint $\bm{X} = \bm{x} \bm{x}^{\mathrm{H}}$, the equations are equivalent to the original quadratic ones.

However, this constraint makes the problem non-convex. In the semi-definite relaxation, we change the constraint to $\bm{X} \succeq \bm{x} \bm{x}^{\mathrm{H}}$, and require $\bm{X}$ being symmetric and positive semi-definite, making the problem convex \cite{Boyd1996_sdp}. Equivalently, the relaxed constraint can be rewritten as
\begin{align}
    \begin{bmatrix}
        \bm{X} & \bm{x} \\ \bm{x}^{\mathrm{H}} & 1
    \end{bmatrix} \succeq 0.
\end{align}
Thus, we end up with the problem \eqref{eq:SDR_MMF} with formulas for~$f_i$, $g_j$, and $h_j$.

\subsection{Result extraction}
Let $(\bm{X}^{\star}, \bm{x}^{\star})$ be the solution to the semi-definite relaxed problem \eqref{eq:SDR_MMF}. The solution does not typically satisfy $\bm{X}^{\star} = \bm{x}^{\star} (\bm{x}^{\star})^{\mathrm{H}}$ because the equality was not required in the relaxed optimization. Therefore, it is unlikely that reflection coefficients can directly be computed based on the SDR solution such that they would satisfy the original constraints.

Nevertheless, we can still compute the EEPs corresponding to $(\bm{X}^{\star}, \bm{x}^{\star})$ of all elements to the directions denoted by the index $l$. These fundamental bounds for EEPs can be computed as
\begin{align}
    |\hat{\bar{e}}_i|^2 = f_i(\bm{x}^{\star}, \bm{X}^{\star}) + |\tilde{\bar{e}}_i|^2,
\end{align}
where $|\tilde{\bar{e}}_i|^2$ contains the vectorized target EEPs. Due to the vectorization, the index $i$ refers to both a driven port $n$ and a direction $l$. The vectorization is inverted to obtain the $\ND \times L$ matrix of EEPs. Note that, the bound EEPs can be computed only toward the directions $(\theta_l, \varphi_l)$, whereas the MO result EEPs can be computed for arbitrary directions as the reflection coefficients are known.

Although we use the SDR mainly for computing the bound, we also try to extract a realizable reflection vector from the solution $(\bm{X}^{\star}, \bm{x}^{\star})$. We compute the reflection coefficients matching to $\bm{x}^{\star}$ using~\eqref{eq:bar_rj}. After that, we take the first $\NP$ terms of the vector $\bar{\bm{r}}^{\star}$, and force the constant modulus constraints as
\begin{align}
    r_m^{\star, \mathrm{CMC}} =  \frac{r_m^{\star}}{|r_m^{\star}|}.
\end{align}
The realizable SDR results are obtained using these reflection coefficients for the scatterer ports.


\bibliographystyle{IEEEtran}
\bibliography{IEEEabrv,references}




\begin{IEEEbiography}[{\includegraphics[width=1in,height=1.25in,clip,keepaspectratio]{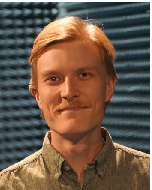}}]{Albert Salmi}
received the B.Sc. and M.Sc. degrees in electrical engineering from Aalto University, Espoo, Finland, in 2020 and 2021, respectively. He has been with the Department of Electronics and Nanoengineering, School of Electrical Engineering, Aalto University, since 2019, where he is currently pursuing a doctoral degree.

His research interests include antenna arrays, optimization, and spherical wave expansion. Recent research has focused on the design of reactively loaded sparse antenna arrays with grating-lobe mitigation.
\end{IEEEbiography}

\begin{IEEEbiography}[{\includegraphics[width=1in,height=1.25in,clip,keepaspectratio]{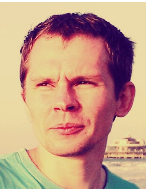}}]{Miloslav Capek}
(M'14, SM'17) received the M.Sc. degree in Electrical Engineering 2009, the Ph.D. degree in 2014, and was appointed a Full Professor in 2023, all from the Czech Technical University in Prague, Czech Republic.
	
He leads the development of the AToM (Antenna Toolbox for Matlab) package. His research interests include electromagnetic theory, electrically small antennas, antenna design, numerical techniques, and optimization. He authored or co-authored over 165~journal and conference papers.

Dr. Capek is the Associate Editor of IET Microwaves, Antennas \& Propagation. He was a regional delegate of EurAAP between 2015 and 2020 and an associate editor of Radioengineering between 2015 and 2018. He received the IEEE Antennas and Propagation Edward E. Altshuler Prize Paper Award~2022 and ESoA (European School of Antennas) Best Teacher Award in~2023.
\end{IEEEbiography}

\begin{IEEEbiography}[{\includegraphics[width=1in,height=1.25in,clip,keepaspectratio]{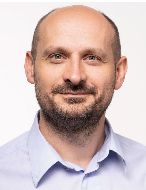}}]{Lukas Jelinek} was born in Czech Republic in 1980. He received his Ph.D. degree from the Czech Technical University in Prague, Czech Republic, in 2006. In 2015 he was appointed Associate Professor at the Department of Electromagnetic Field at the same university.

His research interests include wave propagation in complex media, electromagnetic field theory, metamaterials, numerical techniques, and optimization.
\end{IEEEbiography}

\begin{IEEEbiography}[{\includegraphics[width=1in,height=1.25in,clip,keepaspectratio]{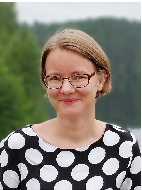}}]{Anu Lehtovuori} received the M.Sc. (Tech.) and Lic.Sc. (Tech.) degrees from the Helsinki University of Technology, Espoo, Finland, in 2000 and 2003, respectively, and the D.Sc. (Tech.) degree from Aalto University, Finland, in 2015, all in electrical engineering. 

She is currently a Senior University Lecturer and Deputy Head of Department with the School of Electrical Engineering, Aalto University, Finland. Her current research interests include multiport antennas, electrically small antennas especially for mobile devices, and antenna-amplifier interaction in antenna systems.
\end{IEEEbiography}

\begin{IEEEbiography}
[{\includegraphics[width=1in,height=1.25in,clip,keepaspectratio]{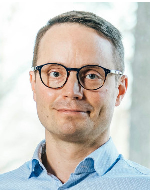}}]{Ville Viikari}
(S’06–A’09–M’09–SM’10) received the Master of Science (Tech.) and Doctor of Science (Tech.) (with distinction) degrees in electrical engineering from the Helsinki University of Technology (TKK), Espoo, Finland, in 2004 and 2007, respectively.

He is currently a Professor and Head of Department with the Aalto University School of Electrical Engineering, Espoo, Finland. From 2001 to 2007, he was with the Radio Laboratory, TKK (now part of Aalto University), where he studied antenna measurement techniques at submillimeter wavelengths and antenna pattern correction techniques. From 2007 to 2012, he was a Research Scientist and a Senior Scientist with the VTT Technical Research Centre, Espoo, Finland, where his research included wireless sensors, RFID, radar applications, MEMS, and microwave sensors. He was appointed an Assistant Professor at Aalto University in 2012. He has authored or co-authored more than 100 journal papers and 100 conference papers. He is an inventor in 16 granted patents. His current research interests include antennas for mobile devices and networks, antenna clusters and coupled arrays, RF-powered devices, and antenna measurement techniques.

Dr. Viikari was a regional delegate of EurAAP 2018-2023. He was the recipient of the Young Researcher Award of the Year 2014, presented by the Finnish Foundation for Technology Promotion and IEEE Sensors Council 2010 Early Career Gold Award.
\end{IEEEbiography}

\end{document}